\documentclass[sigconf]{acmart}
\copyrightyear{2022}
\acmYear{2022}
\setcopyright{rightsretained}
\acmConference[CCS '22]{Proceedings of the 2022 ACM SIGSAC Conference on Computer and Communications Security}{November 7--11, 2022}{Los Angeles, CA, USA}
\acmBooktitle{Proceedings of the 2022 ACM SIGSAC Conference on Computer and Communications Security (CCS '22), November 7--11, 2022, Los Angeles, CA, USA}
\acmDOI{10.1145/3548606.3559381}
\acmISBN{978-1-4503-9450-5/22/11}

\usepackage{etoolbox}
\makeatletter
\patchcmd{\maketitle}{\@copyrightpermission}{
   \begin{minipage}{0.3\columnwidth}
     \href{https://creativecommons.org/licenses/by/4.0/}{\includegraphics[width=0.90\textwidth]{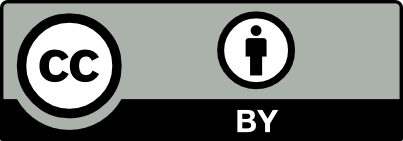}}
   \end{minipage}\hfill
   \begin{minipage}{0.7\columnwidth}
     \href{https://creativecommons.org/licenses/by/4.0/}{This work is licensed under a Creative Commons Attribution International 4.0 License.}
   \end{minipage}

   \vspace{5pt}
}{}{}
\makeatother

\usepackage{xspace}
\usepackage{stmaryrd}
\usepackage{comment}
\usepackage{graphicx}
\usepackage{appendix}
\usepackage{hyperref}

\usepackage{amsfonts}
\usepackage{amsmath}

\usepackage{booktabs}
\usepackage{caption}
\usepackage{tikz}
\usepackage{xcolor}
\usepackage{listings}
\usepackage{multirow}
\usepackage{syntax} 
\usepackage{xparse}
\usepackage{algorithm}
\usepackage{threeparttable}
\usepackage{proof}
\usepackage{enumitem}
\usepackage{tablefootnote}
\usepackage[normalem]{ulem}
\usepackage{array}
\usepackage{pifont}

\newtheorem{example}{Example}

\AtBeginDocument{%
  \providecommand\BibTeX{{%
    \normalfont B\kern-0.5em{\scshape i\kern-0.25em b}\kern-0.8em\TeX}}}

\begin{document}

% \fancyhead{} %removes conference name and paper title from header in each page
\newcommand{\program}{P\xspace}
\newcommand{\query}{Q\xspace}
% \newcommand{\toolname}{{\sc Iustitia}\xspace}
%probable names: OAuthSabre (french), AuthSaber (american), Saber (sword), Cerberus (protector), 

% 1. Cerberus: a greek mythical multi-headed dog that guards the gates of the Underworld to prevent the dead from leaving.

% 2. Saber/Sabre (french): a heavy cavalry sword with a curved blade and a single cutting edge. (OAuthSabre)

% 3. Grogu: A jedi (Protectors for galaxy) character from Star Wars.

% 4. Beskar: a precious metal, which no blaster or lightsaber can penetrate (Mandalorian, Star Wars)

% \newcommand{\toolname}{{OAuthShield}\xspace}
% \newcommand{\toolvariant}{{OAuthShield$^\prime$}\xspace}

% \newcommand{\toolname}{{\textsc{Cerberus}}\xspace}
\newcommand{\toolname}{{\textit{Cerberus}}\xspace}
\newcommand{\toolvariant}{{\textit{Cerberus}$_{eager}$}\xspace}
\newcommand{\oauthlint}{{\textit{OAuthlint}}\xspace}
\newcommand{\skvetter}{{\textit{S3kvetter}}\xspace}
\newcommand{\oauthserver}{{OAuth server}\xspace}

\newcommand{\datasetKnown}{{$\mathtt{DataSet}_K$}\xspace}
\newcommand{\datasetUnknown}{{$\mathtt{DataSet}_U$}\xspace}

\newcommand{\hotspot}{\pi}
\newcommand{\pred}{\phi}

\newcommand{\rosette}{{\sc Rosette}\xspace}
\newcommand{\lang}{\mathcal{L}}

\newcommand\tamjid[1]{{\color{blue}{{#1}}}}
\newcommand\yu[1]{{\color{black}{{#1}}}}
\newcommand\yuan[1]{{\color{cyan}{{#1}}}}
%\newcommand\yu1[1]{{\color{red}{{#1}}}}

% \newcommand\revRR[1]{{\color{blue}{{\textbf{[RR#1]}}}}}
% \newcommand\revA[1]{{\color{blue}{{\textbf{[A#1]}}}}}
% \newcommand\revB[1]{{\color{blue}{{\textbf{[B#1]}}}}}
% \newcommand\revC[1]{{\color{blue}{{\textbf{[C#1]}}}}}
% \newcommand\revD[1]{{\color{blue}{{\textbf{[D#1]}}}}}

% red color text
\newcommand\redtext[1]{{\color{magenta}{{#1}}}}
\newcommand\bluetext[1]{{\color{blue}{{#1}}}}
% red color strikethrough text
\newcommand\redsout{\bgroup\markoverwith{\textcolor{red}{\rule[0.5ex]{2pt}{0.4pt}}}\ULon}

\newcommand{\declStmt}[2]{{#1}\,{#2}}
\newcommand{\letStmt}[2]{\texttt{let}\,\,{#1} = {#2}}
\newcommand{\assumesStmt}[1]{\texttt{assume}\,\,{#1}}
\newcommand{\ensuresStmt}[1]{\texttt{assert}\,\,{#1}}
\newcommand{\spec}{\ensuremath{\psi}}

\lstset{escapeinside={(*}{*)},mathescape}
\lstset{basicstyle=\footnotesize\ttfamily,breaklines=true,numbers=left,stepnumber=1}
\lstset{frame=bottomline}

\newcommand{\irule}[2]%
   {\mkern-2mu\displaystyle\frac{#1}{\vphantom{,}#2}\mkern-2mu}

\newcommand \emptycirc[1][1ex]{\tikz\draw (0,0) circle (#1);} 
\newcommand \halfcirc[1][1ex]{%
  \begin{tikzpicture}
  \draw[fill] (0,0)-- (90:#1) arc (90:270:#1) -- cycle ;
  \draw (0,0) circle (#1);
  \end{tikzpicture}}
\newcommand \fullcirc[1][1ex]{\tikz\fill (0,0) circle (#1);}

\newcommand{\den}[1]{\llbracket#1\rrbracket}
\newcommand{\concat}{\rightarrow}
\newcommand{\negate}{!}
\newcommand{\disjunct}{+}

% \title{\toolname: Efficient Security Checking for OAuth Service Provider Implementations}
%\title{\toolname: Query-driven Security Checking for OAuth Service Provider Implementations at Scale}
\title{\toolname: Query-driven Scalable Vulnerability Detection in OAuth Service Provider Implementations}

% \author{}
\author{Tamjid Al Rahat}
\affiliation{%
  \institution{University of California, Los Angeles}
  \country{}}
\email{tamjid@g.ucla.edu}

\author{Yu Feng}
\affiliation{%
  \institution{University of California, Santa Barbara}
  \country{}}
\email{yufeng@cs.ucsb.edu}

\author{Yuan Tian}
\affiliation{%
  \institution{University of California, Los Angeles}
  \country{}}
\email{yuant@ucla.edu}


%-------------------------------------------------------------------------------
\begin{abstract}
%-------------------------------------------------------------------------------
OAuth protocols have been widely adopted to simplify
user authentication and service authorization for third-party applications. However, little effort has been devoted to automatically checking the security of the libraries that service providers widely use.
In this paper, we formalize the OAuth specifications and security best practices, and design 
\toolname, an automated static analyzer, to find logical flaws and identify vulnerabilities in the implementation of OAuth service provider libraries. 
To efficiently detect security violations in a large codebase of service provider implementation, \toolname employs a query-driven algorithm for answering queries about OAuth specifications. 
We demonstrate the effectiveness of \toolname by evaluating it on datasets of popular
OAuth libraries with millions of downloads. Among these high-profile libraries, \toolname has identified 47 vulnerabilities from ten classes
of logical flaws, 24 of which were previously unknown. We got acknowledged by the developers of eight libraries and had three accepted CVEs.
\end{abstract}


\maketitle

\fancypagestyle{firstpage}
{
    \fancyhead[]{}
    \fancyhead[C]{\textcolor{red}{Appeared in ACM SIGSAC Conference on Computer and Communications Security (CCS), November 2022. Please cite the conference version of the paper.}} 
}
\thispagestyle{firstpage}

\section{Introduction}~\label{sec:intro}
OAuth is widely used for authorizations across different software services. It defines a process for end-users (resource owners) to grant a third-party website/application (also noted as relying party or client applications~\cite{rfc6749}) access to their private resources stored on a service provider, without sharing their passwords for the service providers with the client applications. As a multi-party protocol, the security of OAuth depends on the service providers (also noted as authorization server or \textit{\oauthserver}~\cite{rfc6749}), the client applications, and the resource owners. In particular, vulnerabilities in \oauthserver implementations could lead to severe consequences because that would impact all the client applications (along with all the resource owners of the applications) that work with the service provider. For example, a vulnerability discovered in 2019 in the Microsoft \oauthserver's redirect URI validation mechanism allowed the attackers to take over Microsoft Azure Accounts~\cite{news-microsoft}. Also, in 2018, a vulnerability in Facebook \oauthserver allowed the attackers to steal access tokens (issued by authorization server) of almost 50 million users~\cite{news-facebook}.

%The reason why these popular service providers such as Microsoft and Facebook still make severe mistakes is that securing OAuth server-side implementation is challenging. Firstly, OAuth protocol is complex and its security measures often vary with different client's attributes. For example, security measures for native applications are different from web applications. Secondly, though the official specification~\cite{rfc6749} is provided for OAuth, developers inadvertently miss or incorrectly implement the security checks required for handling the OAuth server requests. Many attacks in the form of token hijack or authorization code injection have been observed in the wild with the OAuth servers that claims to follow the specification. Thirdly, existing methods~\cite{sun2012devil,chen2014oauth} mostly analyze the security of OAuth servers by intercepting the network traffic during OAuth requests--while considering the server as blackbox. They often can not identify the logical mistakes in the implementation systematically and can not help developers to detect and fix the vulnerabilities efficiently. 

Even well-known service providers such as Microsoft and Facebook still make severe mistakes in their authorization server implementation -- which leads to a pressing question, \textit{how to vet OAuth service provider implementation for security?} Since OAuth is very high-profile and critical for user security and privacy, many previous works have been on OAuth security. However, as far as we know, there is no automatic tool to identify the security-sensitive logical flaws in the service provider implementation. 

Researchers have devoted significant effort to investigating OAuth security by analyzing the protocol~\cite{yang2013security} and proposing formal verification tools~\cite{fett2016comprehensive,pai2011formal}. However, these tools analyze the security at the protocol level and do not consider the diversified implementation details of \oauthserver. Chen et al.~\cite{chen2014oauth} perform the first in-depth study on the security issues of OAuth by running a manual analysis of mobile apps and monitoring the network traffic. Later on, researchers develop semi-automated tools~\cite{bai2013authscan, zhou2014ssoscan} to report various security issues with OAuth client applications. However, these analyses often need an extensive manual setup and do not scale for large programs. 
% \redtext{RR1: Clearly state what is the technical novelty of this work in comparison with the related work, namely [55] and [69]}
Most recently, 
% researchers proposed automatic tools for checking the security properties in relying party programs, which are smaller than service provider programs. For example,
\skvetter~\cite{yang2018vetting} performs symbolic execution, and \oauthlint~\cite{DBLP:conf/kbse/RahatFT19} leverages whole-program data-flow analysis to check client-side properties. As we show later in the evaluation, these approaches are not applicable for checking \oauthserver properties for the following reasons: first, symbolic execution suffers from path explosion~\cite{xie2009fitness} and whole-program analysis also does not scale well, especially for large codebases like \oauthserver programs. Second, data-flow predicates alone are not expressive enough to describe crucial properties that require control-flow predicates.

In addition, most existing works assume that the OAuth servers are securely implemented, which clearly is  not the case according to our study in this paper.
As a result, it is urgent to help developers check the security of the \oauthserver implementations. Unfortunately, as far as we know, there is no systematic security study of \oauthserver implementations. Most previous works in OAuth security either focus on client applications or report security issues on an ad hoc basis.

We compare the representative OAuth tools from previous work using five relevant criteria: (1) \textit{Automated:} ability to identify vulnerabilities automatically, (2) \textit{Coverage:} the ability to cover and reason about the behavior of OAuth implementation, (3) \textit{Server-side flaws:} the ability to analyze logical flaws of service provider implementations, (4) \textit{Extensibility:} the ability to provide an interface to define and check new security properties, (5) \textit{Scalability:} scalable for extensive programs. We summarize the comparison in Table~\ref{tab:tool-comparison}, which clearly shows the gap.
%\yuan{Please compare the following two flows, which one do you prefer?}

%Researchers have investigated OAuth at the protocol level and propose formal verification tools~\cite{fett2016comprehensive,pai2011formal}. However, these tools only work at the protocol level and do not consider the diversified implementation details of OAuth service providers. Later on, researchers develop semi-automated tools~\cite{bai2013authscan, zhou2014ssoscan} to report various security issues with OAuth relying party implementations. These analyses often need heavy manual setup and do not scale. Most recently, researchers propose automatic tools for checking the implementation mistakes by relying parties~\cite{DBLP:conf/kbse/RahatFT19,yang2018vetting}, assuming the service providers are securely implemented, which is clearly not the case according to our study in this paper. As far as we know, there are only very few adhoc reports of service providers implementation mistakes such as Chen et al. ~\cite{chen2014oauth}. However, since the paper~\cite{chen2014oauth} focuses on mobile app analysis, the work cannot detect the logical flaws in the server-to-server flows. As a result, the paper still mostly identifies relying party implementation issues rather than service provider implementation issues. In addition, it is based on manual analysis, thus do not scale. \yuan{Shall we delete the last three sentences? Do they read too defensive?}

To bridge the gap, we propose systematically checking the security of \oauthserver implementations of authorization flows. We start by investigating popular open-source \oauthserver libraries. We find many developers use these libraries to implement their service provider applications instead of starting from scratch due to their complexity. For example, Node OAuth2 Server~\cite{lib-node-oauth2-server} library is used by more than 2,400 repositories on Github. Since developers widely use these open-source libraries  to integrate OAuth in their implementations, checking the security of \oauthserver implementations in these libraries also provides us insights about many implementations built on top of these libraries. 

%build a tool that extract the states and logic flows in the servers and verify the security properties. 

\begin{table}[b]
    	\centering
    	\footnotesize
    	\caption{Comparison with existing OAuth analysis tools.}
    
    	\begin{tabular}
    		{p{0.15\textwidth}	% Tool
    			p{0.04\textwidth}		% Automated
    			p{0.04\textwidth}		% Correctness
    			p{0.04\textwidth}		% Server-side (SP) logic
    			p{0.04\textwidth}		% Extensibility (Discover new bugs)
    			p{0.04\textwidth}} %scalability
    		
    		\toprule
    		\textbf{Tool} & \textbf{Auto.} & \textbf{Covg.} & \textbf{Serv.} & \textbf{Ext.} & \textbf{Scal.}\\
    		\midrule
    		
    		AuthScan~\cite{bai2013authscan} & \halfcirc & \halfcirc & \emptycirc & \emptycirc & \emptycirc\\
    		
    		SSOScan~\cite{zhou2014ssoscan} & \fullcirc & \halfcirc & \emptycirc & \emptycirc & \emptycirc\\
    		
    		 OAuthTester~\cite{yang2016model} & \fullcirc & \emptycirc & \emptycirc & \emptycirc & \emptycirc\\
    		
    		S3kVetter~\cite{yang2018vetting} & \fullcirc & \fullcirc & \emptycirc & \fullcirc & \emptycirc\\
    		
    		OAuthLint~\cite{DBLP:conf/kbse/RahatFT19} & \fullcirc & \fullcirc & \emptycirc & \halfcirc & \emptycirc\\
    	%	Chen \emph{et~al.}~\cite{chen2014oauth} & \emptycirc & \tablefootnote{Chen et al.~\cite{chen2014oauth} is based on manual analysis, so there is no tool available.}  \halfcirc & \halfcirc \tablefootnote{Chen et al.~\cite{chen2014oauth} focuses on analyzing mobile apps, as a result, the work mostly reports implementation issues in relying parties rather than service providers. The detailed comparisons are presented in the related work section.}& \emptycirc \\
    		\midrule
    		\toolname & \fullcirc & \fullcirc & \fullcirc & \fullcirc & \fullcirc\\
    		
    		\bottomrule
    	\end{tabular}
    	
    	\caption*{\fullcirc[0.75ex] Full support \halfcirc[0.75ex] Partial support \emptycirc[0.75ex] No support. \\\hspace{\textwidth} \textbf{Auto.}: Automated, \textbf{Covg.}: Coverage, \textbf{Serv.}: Server-side flaws, \textbf{Ext.}: Extensibility,  \textbf{Scal.}: Scalability}
    	
    	\vspace{-5mm}
\label{tab:tool-comparison}
\end{table}

\noindent\textbf{Challenges}: 
However, developing an automatic tool for finding logical flaws of \oauthserver implementations is quite challenging due to the following reasons:
\setlist{nolistsep}
\begin{itemize}[noitemsep]
    \item \emph{Expressiveness}. The original specifications for OAuth protocols are written in plain English, which are difficult to be turned into checkable invariants consumed by the analyzers. Consequently, properties for implementing a secure and effective server-side implementation for authorization using OAuth are not well-defined.
    \item \emph{Scalability}. Implementing the protocols on \oauthserver is typically complex and large and often relies on many third-party libraries. Therefore, as shown  in Sec.~\ref{sec:eval}, a naive whole-program analysis will not scale well.
    \item \emph{Generality}. Checking security properties in \oauthserver implementations is not straightforward as they depend on various factors such as the client's platform and types of grants (i.e., authorization flows) used by the client during the authorization process. Thus, it is non-trivial to devise a framework that unifies all properties enforced by the OAuth specifications.

\end{itemize}

% \noindent\textbf{Goals:}
% In this paper, we build \toolname, a tool that automatically verify the security properties in those the server-side implementations of popular OAuth libraries. We have the following goals for the tool:
% \begin{itemize}
%     \item Accuracy. The tool needs to be accurate in identifying vulnerabilities. 
%     \item Efficiency. The tool should be efficient when analyzing the server-side implementations. 
%     \item Scalablity. It should be easy to adapt the tool for new libraries and implementations. 
% \end{itemize}

\noindent\textbf{Our solution}: 
% \tamjid{IMPORTANT: This paragraph seems to confuse the reader as its quickly moving into a lot of technical details about automaton and sub-callgraph construction. Can we be more abstract here and refer the reader to more technical details in Sec 3 and 4?} 
% \yuan{Should also mention the new dynamic analysis.}
To address the challenges mentioned above, we design and implement \toolname, a scalable and automated static analysis tool for detecting logical flaws in large-scale \oauthserver programs written in server-side languages like Java and Javascript.
In particular, we first design a domain-specific language (DSL) to enable security analysts to express the security properties of \oauthserver that are recommended by OAuth specification ~\cite{rfc6749} and OAuth current security best practices~\cite{oauth-best-practice}.

% \tamjid{that are responsible for causing the commonly observed attacks (e.g., redirect URI manipulation, access token injection, etc.) on the authorization server. These properties are also required to be implemented as per OAuth 2.0 specifications~\cite{rfc6749} and OAuth security best practices~\cite{oauth-best-practice}. In total, we identify ten security properties for the OAuth authorization server, five of which were not studied before (details are in Table ~\ref{tab:list-of-properties})}.

Second, given an OAuth property $\query$ encoded in our query language, 
we represent both the desired property $\query$ and the \oauthserver programs as \emph{System Dependence Graphs} (SDGs). Therefore, \toolname converts the problem of checking OAuth properties into a graph query problem, which an off-the-shelf Datalog engine~\cite{souffle} can answer. Note that our graph representation aims to strike a good balance between expressiveness and scalability by capturing temporal sequencing of API calls, data flows between arguments and returns of a procedure, data flows between various program objects, etc. 

However, there is a steep trade-off between the precision of an SDG and the cost of constructing it. For example, SDGs 
constructed using a context-insensitive pointer analysis tend to grossly overapproximate the targets of virtual method calls, which leads to unacceptable false alarms. On the other hand,
more precise SDGs obtained using context-sensitive
pointer analysis can take hours to construct.  Currently, analyses that rely on system dependence graph information must implement their own ad hoc analysis~\cite{he2015vetting, egele2013empirical} to answer application-specific queries. To mitigate this challenge and only reason about program fragments relevant to the query, we introduce a query-driven approach based on automata theory. Since the original system dependence graph is obtained by stitching all control flow graphs from the methods, our key intuition is to keep track of the methods  relevant to the OAuth property. In particular, we define the OAuth request endpoint, which is the precondition of the OAuth property, specifying the entry and exit points of the relevant code snippet. After that, we leverage the request endpoints to pinpoint a sub-callgraph between the request endpoints. Since the sub-callgraph is typically small, its corresponding system dependence graph will also be small.

\noindent\textbf{Findings}:
%With our tool, we investigate the implementations of two popular multi-party protocols: OAuth 2.0 and OpenID Connect.
We evaluate our tool with the ten most popular libraries that use the standard OAuth protocol for implementing \oauthserver and find pervasive vulnerabilities in these popular libraries. All the ten popular libraries we studied have at least one security-critical property violation (i.e., vulnerability). In total, we identified 47 vulnerabilities, 24 of which were previously unknown. We study ten security-critical logical flaws in \oauthserver, five of which (P4, P5, P6, P7, and P9 in Table~\ref{tab:list-of-properties}) are studied by us for the first time. Violation of these properties may lead to severe attacks that are commonly observed on OAuth servers. We got acknowledged by the developers of eight libraries, among which five libraries, including Spring Auth Server~\cite{lib-spring-authorization-server} and Node OAuth2 Server~\cite{lib-node-oauth2-server}), immediately took actions to fix the vulnerabilities. Three classes of new vulnerabilities in these libraries also lead us to have new CVE entries (CVE-2020-26877, CVE-2020-26937, and CVE-2020-26938).
% \footnote{The CVEs are currently marked as ``Reserved'' and will be published after we make the vulnerabilities public}. 
These vulnerabilities can lead to account breaches and confidential information leakage for both clients and resource owners. Eight libraries immediately acknowledged our findings; five have fixed the issues after we reported them, and the rest of the libraries are currently taking action. Finally, \toolname is effective in that it significantly outperforms \oauthlint, a state-of-the-art checker~\cite{DBLP:conf/kbse/RahatFT19} in terms of running time ($>20\times$ faster) and expressiveness.  

\noindent\textbf{Contributions.}
In summary, we make the following contributions:
\setlist{nolistsep}
\begin{itemize}[noitemsep]
    \item We identify and formalize the security properties required to implement the OAuth protocol on the service provider from the standard specifications and security best practices for OAuth.
    \item We, for the first time, design and implement an efficient automated tool, \toolname, to find logical flaws in \oauthserver implementation based on
    % correctly implements security properties recommended by 
    the standard specifications and security best practices.
    \item Our analysis finds that many open-source libraries for \oauthserver make critical security mistakes due to omitting or incorrectly implementing the security properties.
    \item We show that our query-driven scalable analysis is $25\times$ faster in identifying the security property violations when compared with the eager analysis approach over the entire program.
\end{itemize}

% \begin{table}
% \centering
% \begin{tabular}{|l|c|c|}
% \hline
% \multicolumn{1}{|c|}{\textbf{Analysis Method}}
% & \multicolumn{1}{|c|}{\textbf{Scope}}
% & \multicolumn{1}{|c|}{\textbf{Reference}}
% \\ \hline
% Formal protocol analysis
% & client+server
% & \cite{fett2016comprehensive,pai2011formal}
% \\ \hline
% Manual traffic analysis
% & client+server
% & \cite{chen2014oauth,gibbons2014security,li2016analysing}
% \\ \hline
% Fuzzing tools
% & client
% & \cite{yang2016model,mainka2017sok}
% \\ \hline
% Automated static analysis
% & client
% & \cite{yang2018vetting, DBLP:conf/kbse/RahatFT19}
% \\ \hline
% \end{tabular}

% \caption{Existing works and their scope to analyze OAuth security.}
% \label{tab:list-oauth-papers}

% \end{table}

% The rest of the paper is organized as follows: 
% \yuan{Todo: we can finish this paragraph after other parts are stable.}

\section{Background}~\label{sec:background}
This section describes the widely-used authorization protocol OAuth and the communication between different entities during the authorization process. %Although OAuth has two major versions: OAuth 1.0a and OAuth 2.0, the specifications of both protocols are completely different, and OAuth 2.0 is not backward compatible to OAuth 1.0, thereby cannot be used together. Since OAuth 1.0 has severe security implications and the API is too complex to implement for many clients, OAuth 2.0 is nowadays the most popular and widely used form of OAuth. 
% \yuan{I believe major service providers all stopped supporting OAuth 1.0, please check it out and add references to support the argument.} 
In this paper, we focus on OAuth 2.0, which is the current industry-standard protocol for authorization, and throughout the paper, we refer to OAuth 2.0 when we say OAuth. 
%\subsection{OAuth 2.0}
OAuth 2.0 is an open standard authorization protocol where users delegate client applications (relying parties) to access their information hosted on the service providers without giving away their passwords. %OAuth was originally created in response to the direct authentication pattern, where users are required to present their username and password to the client apps so that the apps can access user's information. 
%To avoid sharing user's sensitive credentials to client's server, a federated identity was created for single sign-on (SSO). In this scenario, resource owner communicates with the authorization server of the service provider (identity provider), and the authorization server generates a cryptographically signed token which it sends to the client's apps to authenticate the user. As long as the trust between the client and service provider holds, client can also use the token to obtain access to the user's protected resources with limited scope set by the user. However, to obtain a token, client uses an authorization grant which is a credential representing the user's authorization to access the protected resources.
% \begin{figure}[!t]
% \centering
%   \includegraphics[scale=0.38]{figures/implicit-flow.pdf}
%   \caption{Implicit grant flow for OAuth 2.0.}
%   \label{fig:implicit-grant}
% \end{figure}
\begin{figure}[!t]
\centering
  \includegraphics[scale=0.4]{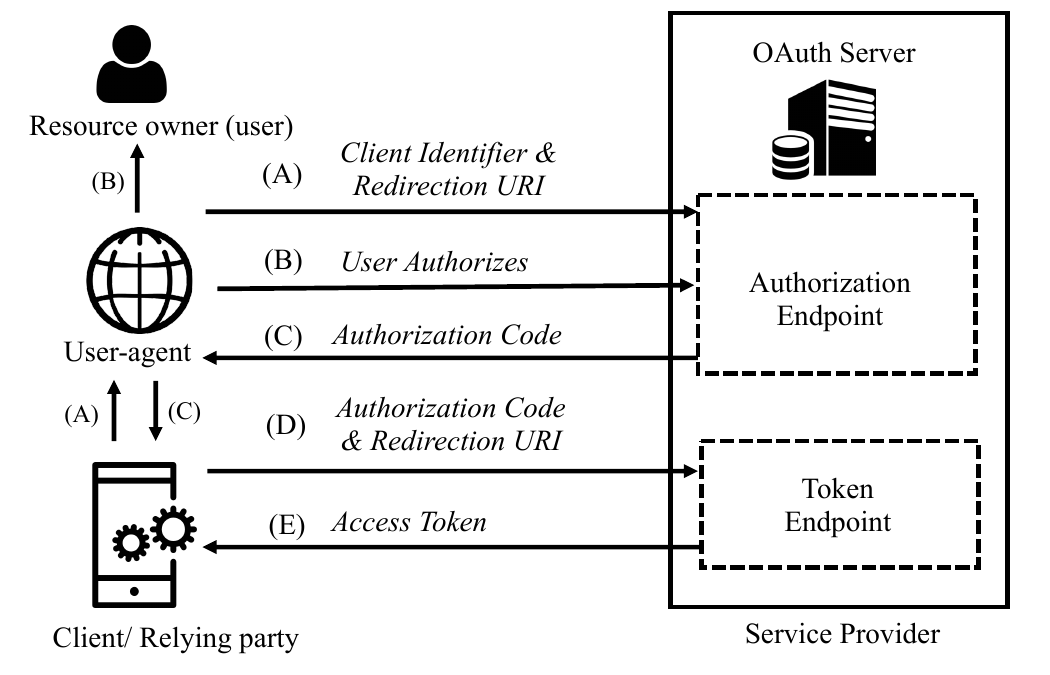}
  \caption{Authorization code grant flow for OAuth.}
  \label{fig:auth-code-grant}
\vspace{-6mm}
\end{figure}
OAuth specification defines four types of grants--(1) authorization code grant, (2) implicit grant, (3) resource owner password credentials grant, (4) client credentials grant. Two grants are most widely used in practice -- authorization code grant and implicit grant. In the following, we explain the details of these two grants.\\  %Among all grants, authorization code grant is the most recommended grant by the OAuth specification as it provides better security benefits than the others. Implicit grant is a simplified authorization code flow designed for the in-browser applications (e.g., single-page javascript apps) that cannot securely store client's credentials. Resource owner password credentials grant requires the client to have direct access to the resource owner password credentials, and it is used when there is high degree of trust between the client and the resource owner. Finally, client credentials grant is used when client itself acts as the resource owner or client has special authorization previously arranged with the authorization server. Since resource owner password credentials and client credentials grants are mostly used by confidential clients in special scenarios, in this paper we focus on the two most popular grants: authorization code and implicit grant, both of which covers a wide range of web and mobile applications that use OAuth.
\noindent
\textbf{Implicit Grant.} Implicit grant is the simplest grant type. It has two steps. 
% Fig.~\ref{fig:implicit-grant} illustrates the implicit grant flow. 
First, the user is redirected to the \oauthserver to grant the relying party (client) access to their protected resources. After the user grants permission, the server redirects the user back to the relying party along with an access token. The relying party can then use this access token to request the user’s protected resource from the \oauthserver. \\
\noindent
\textbf{Authorization Code Grant.} Authorization code grant %is the most common OAuth grant type that 
can be used by both web apps and native apps to obtain the access token. The flow of authorization code grant is illustrated in Fig.~\ref{fig:auth-code-grant}. The authorization code grant augments the implicit grant by adding a step for authenticating the relying party (client). After user grants permission to the relying party, \oauthserver redirects the user back to the relying party. Instead of providing the access token directly to the relying party, the server sends an authorization code this time. Then the relying party can use the authorization code to exchange for the access token by making a new request at the token endpoint. To get the access token, the relying party needs to include its identity in this request so that the server can verify if the authorization code is granted to the same party.

Although the authorization code grant provides better security benefits than the other grants, it is still vulnerable to code interception attacks, specially for public clients (e.g., native desktop apps), where the attackers intercept the authorization code returned from the authorization endpoint (step C in Fig.~\ref{fig:auth-code-grant}) and obtain the access token by exchanging the code at the token endpoint (step D). %This kind of attack commonly appears in client applications that use inter-application communication within the client's operating system. For example, in Android, a malicious application can register itself as a handler for the custom URI scheme in addition to the benign OAuth app and thereby intercept the authorization code returned from the authorization server. The attacker then can use the authorization code to obtain the access token from the token request endpoint. However, to obtain the token, the client also needs to authenticate themselves at the token endpoint by presenting the client identifier and client secret. Public clients such as mobile applications cannot confidentially store the client secrets, and attackers might be able to easily access the secret by doing a simple revere engineering. 
To mitigate the risk of code interception attack, OAuth specification introduces an extension of the authorization code grant called Proof Key for Code Exchange (PKCE)~\cite{rfc7636} and requires all OAuth servers to support PKCE for public clients. It is worth mentioning that, although PKCE was originally designed to protect public clients, it is recommended~\cite{oauth-best-practice} to use PKCE for all kinds of clients, including web applications.

% \begin{figure}[!t]
% \centering
%   \includegraphics[scale=0.4]{figures/pkce-flow.pdf}

%   \caption{Proof Key for Code Exchange (PKCE) -- an extension of authorization code grant to provide more security benefits for client applications.}
%   \label{fig:pkce-flow}
% \end{figure}
\noindent
\textbf{PKCE.} Since public clients cannot maintain confidentiality and cannot securely store the client's secret, using PKCE allows \oauthserver to authenticate clients without the secret key. PKCE utilizes a dynamically created cryptographic random key called \textit{code verifier}. A unique \textit{code verifier} is generated by the client for every authorization request. The transformed value of the \textit{code verifier}, called \textit{code challenge}, is sent to the \oauthserver to obtain the authorization code. When a client makes a new request at the token endpoint to obtain the access token, it also sends the \textit{code verifier} along with the authorization code it received from the previous request. To validate the proof of possession of the \textit{code verifier} by the client, \oauthserver transforms the \textit{code verifier} and validates it with the previously received \textit{code challenge}. 
% For example, for the code challenge method of `S256', the server transforms the \textit{code verifier} by using \texttt{Base64UrlEncode(Sha256))} and compares if the transformed value is equal to the \textit{code challenge}, which it received in the previous state of the flow. 
This approach helps to mitigate the authorization code injection attack as an intercepted authorization code from the authorization endpoint cannot be exchanged for an access token without the one-time key of \textit{code verifier}.
\section{Overview}~\label{sec:overview}
This section briefly explains how our tool detects vulnerabilities in \oauthserver implementations using a motivating example. In what follows, we first describe the threat model our system, and then with a real-world example, we explain how insufficient security checks in the \oauthserver allow malicious clients to steal sensitive OAuth credentials and how our tool is designed to detect such logical flaws.

% \begin{figure*}[!t]
%   \includegraphics[width=\linewidth]{figures/tool_overview.png}
%   \caption{Overview of the tool}
%   \label{fig:overview}
% \end{figure*}

\noindent\textbf{Threat Model:}~\label{sec:threat}
%In this paper, we aim to detect vulnerabilities in the OAuth-server (Service Provider) at the implementation level. We assume attackers can be any client app, client server or even a resource owner who can access the OAuth-server APIs by making HTTP request to different endpoint provided by the server. In other words, OAuth-server behaves as a black-box to the attackers. Attackers cannot modify any program behavior of the source code and the underlying logic for handling OAuth-requests by the OAuth-server may be unknown to the attacker. However, attacker may have full access to the client application and make API requests on behalf of a benign client. Thus, attackers may intercept the authorization request made by client to perform CSRF attacks and steal the authorization code. Attackers may also launch phishing attacks to obtain authorization from resource owners and thereby, steal the authorization code or access token. At token endpoint, attackers may launch a code-replay attack to hijack the token of benign resource owners. Moreover, at resource endpoint, attackers may misuse a stolen token to impersonate resource owners and get access to their protected resources.  
We aim to detect vulnerabilities in the \oauthserver at the implementation level. We assume attackers can be a malicious relying party or a malicious user (resource owner) who interacts with the victim \oauthserver, also known as service provider. We assume the attackers cannot directly modify the source code or logic of the service provider but can initiate attacks by sending requests to their server. The relying party attackers control their own malicious relying party apps. For example, the relying party attackers might send malicious requests to the victim service provider to access the user's information without the user's approval. The resource owner attackers use their own devices to communicate with the benign service provider to log in on behalf of the victim user. %they have access to the malicious relying party app.

\noindent\textbf{Vulnerable implementation.}~\label{subsec:vul-implementation}
The industry-standard authorization protocol of OAuth %is designed to allow third-party client applications (e.g., mobile apps) to obtain access to an HTTP service (e.g., API access) on behalf of a user. The protocol 
is well designed for access delegation, but a wrong implementation or incorrect usage can have a colossal impact. %During the authorization process, the relying party, depending on the grant they use, gets an authorization code or access token with specific permissions to take actions on behalf of the user to whom the authorization code or token belongs. This highly privileged access token enables attackers to control the account. %, in most cases, equate to having the username and password for the account itself. %Also, if a malicious attacker steals such highly privileged token, it will, in some cases, allow the attacker to bypass the security components like 2FA. 
During the authorization process, client gets an access token with specific permissions to take actions on behalf of the user which can even be used to control user's account. When responding to an authorization request, the \oauthserver passes the authorization code or tokens to the client application using a redirect URI, which describes the destination where the code or tokens are passed. The client application sets an allowed list of trusted URIs to receive the OAuth tokens during the registration process. However, many OAuth servers do not appropriately validate redirect URI, leading to the possibility of passing the tokens to a malicious URI under the attackers' control. In recent years, many attacks exploiting the redirect URI have been observed in the OAuth servers, including the popular ones like Microsoft~\cite{news-microsoft} and Twitter~\cite{twitter-attack}. %The BlackDirect vulnerability, discovered in 2019, allowed the attackers to exploit the $redirect\_uri$ to takeover Microsoft and Azure accounts. 
Similar attacks have been observed in the open-source OAuth servers, as they also utilize the same OAuth protocol.

For example, ApiFest~\cite{apifest-site} is a popular open-source  \oauthserver implementation that uses OAuth protocol to provide a secure API management service. We identify a new vulnerability of incorrect redirect URI validation in this library (CVE-2020-26877). 
% This vulnerability allows the attacker to get the user's data in the service provider without the user's consent. 
%The library supports a customized grant to issue access tokens upon successful authorization process. 
Fig.~\ref{fig:code-example-motivating} demonstrates a simplified implementation of their authorization request endpoint. 
%The customized grant is very similar to the original authorization code grant shown in Figure~\ref{fig:auth-code-grant}. 
ApiFest server makes mistakes in the critical step of checking the redirect URI submitted by the client, which allows an attacker to obtain the authorization code by using a maliciously crafted redirect URI during an authorization request. In particular, this implementation makes two severe security mistakes at this step (line 6-9 in Fig.~\ref{fig:code-example-motivating}): (1) it does not check whether the redirect URI is registered to the corresponding client, (2) it omits the required format checking for the redirect URI. Since the server does not match the submitted redirect URI with the client's registered URI, it allows the attacker to craft the redirect URI parameter with attacker's own redirect URI and thereby, steal the authorization code of a legitimate client through the redirected URI. Additionally, attackers can also leverage open redirectors of the user agent to steal the authorization code. These mistakes also violate the standard OAuth specification as described in RFC-6749~\cite{rfc6749}. 

To detect such security vulnerabilities caused by incorrect implementation of logical properties for OAuth, in this paper, we design and implement an automated and scalable tool, called \toolname, to analyze the large codebase that implements \oauthserver. We first identify the security-sensitive properties based on the standard OAuth specification~\cite{rfc6749} and security best practices~\cite{oauth-best-practice, rfc6819}. Then, we meticulously design a query language to formally express the properties so that developers can easily define them and they are understandable by the analysis tool as well. \toolname then represents \oauthserver programs at the statement level using system dependence graphs (SDGs), while maintaining the control and data flow relationship between the statements. However, as \oauthserver programs can be huge, running analyses on the statement-level representation for the whole program do not scale well. To overcome the scalability challenge, \toolname automatically identifies the program component corresponding to OAuth query. Finally, after \toolname pin-points its scope in the OAuth relevant implementation in the program, it executes the query to identify the violation of properties that might expose the server to security attacks.

\begin{figure}[!t]
%\lstset{style=codestyle}
\begin{lstlisting}[xleftmargin=2em,escapeinside={(*}{*)}]
public void (*\bfseries authRequest*)(HttpRequest request, HttpResponse response) {
    String clientId = request.client_id;
    Client client = (*\bfseries getClient*)(clientId);
    String redirectUri = (*\bfseries request.redirect_uri*);
    [...]
    //incorrect Redirect URI validation
    (*\bfseries if (redirectUri == null)*) {
        redirectUri = (*\bfseries client.redirect_uri*);
    }
    [...]
    AuthCode authCode = (*\bfseries generateCode*)(clientId, scope, ...);
    authCode.setClient(clientId);
    authCode.setRedirectUri(redirectUri);
    DB.storeAuthCode(authCode);
    [...]
    QueryStringEncoder encUri = new QueryStringEncoder((*\bfseries redirectUri*));
    encUri.addParam("code", (*\bfseries authCode*).getCode());
    //redirects with auth code
    response.(*\bfseries sendRedirect*)(encUri.toString());
}
\end{lstlisting}
\caption{Code example for an authorization endpoint that is vulnerable to the redirect URI manipulation during the authorization flow.}
\label{fig:code-example-motivating}
\vspace{-7mm}
\end{figure}
\section{System Design}
~\label{sec:system}
In this section, we discuss the design and implementation of \toolname,
our end-to-end static analysis tool for systematically checking the security issues of \oauthserver implementation. %with respect to OAuth 2.0 specifications.

%\subsection{Overview}
% Given an OAuth server implementation $P$ as well as an OAuth property $\query$ that is expected to hold in $P$, a naive approach is to perform a whole-program model checking over $P$ to search for counter-examples that violate the desired property $\query$. As we later show in the evaluation, the naive approach will not scale to a large codebase. To address this challenge, \toolname introduces a demand-driven algorithm to reason about a small fraction of $P$ that are relevant to the property $\query$.
\begin{figure}[!t]
\centering
  \includegraphics[width=\linewidth]{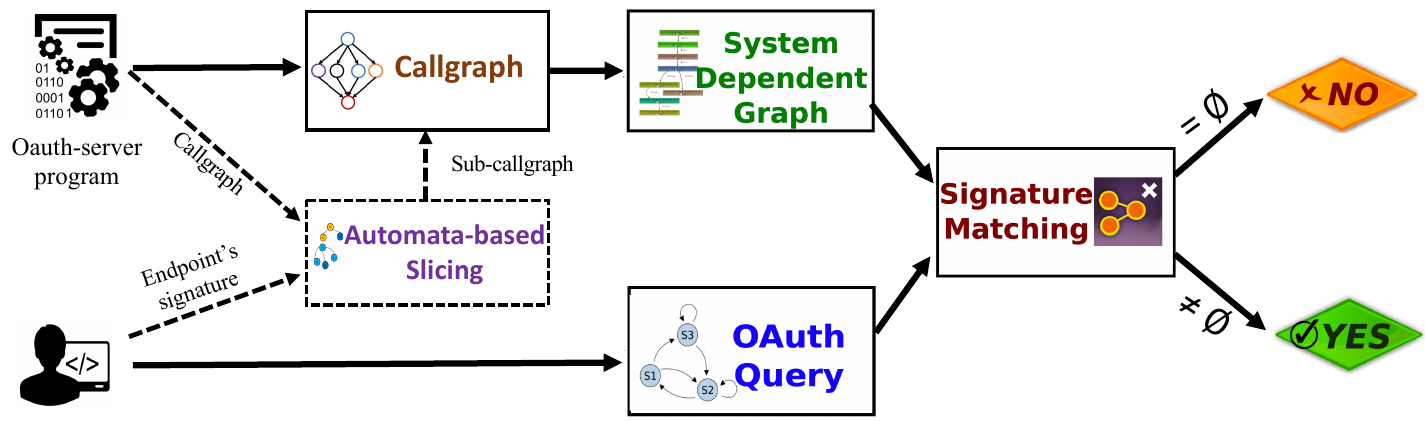}
  \caption{Overview of \toolname. Here, Automata-based slicing is an optimization step discussed in Sec.~\ref{sec:impl}.}
  \label{fig:overview}
\vspace{-6mm}
\end{figure}

\begin{figure*}[!t]
\centering
  \includegraphics[scale=0.44]{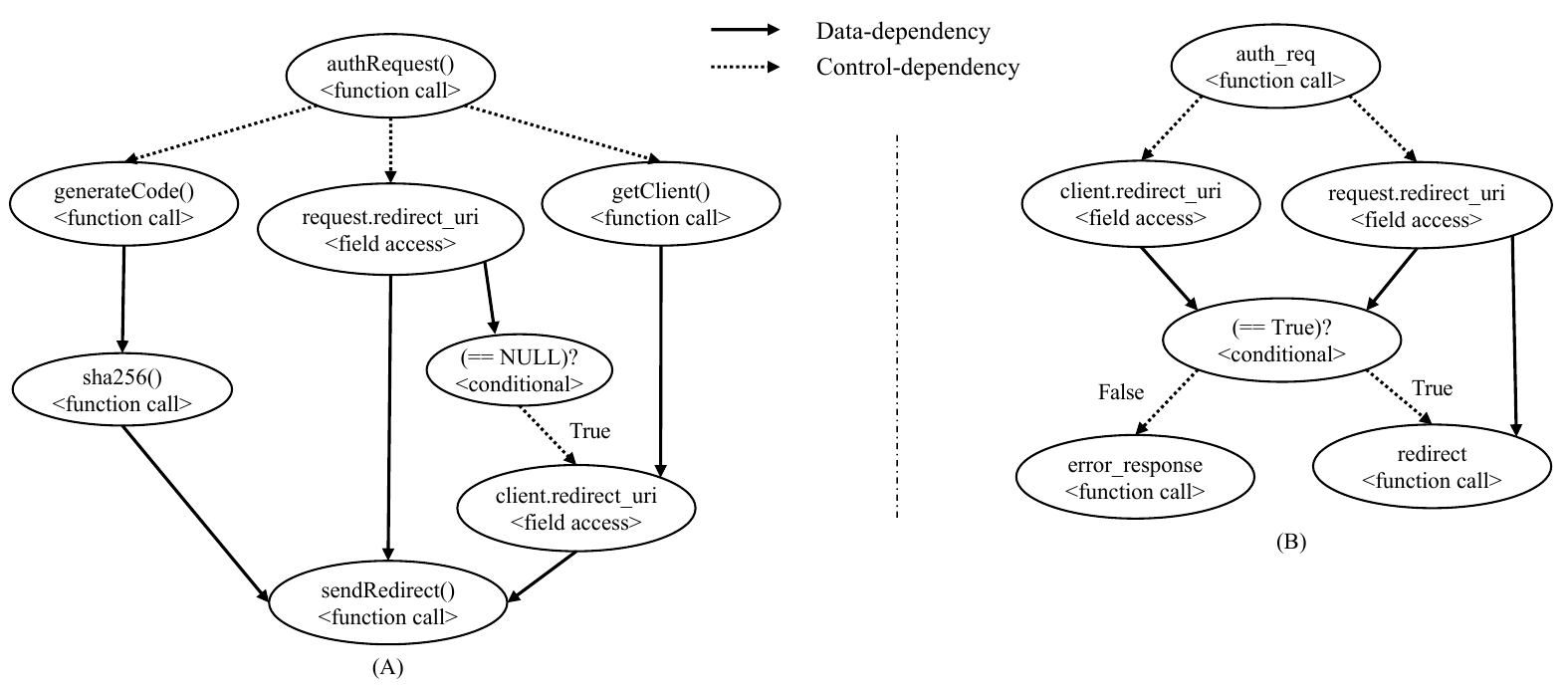}
  \caption{(A) System Dependence Graph (SDG) representation for the code example in Fig.~\ref{fig:code-example-motivating} and (B) graphical representation of the query for the Redirect URI property (P1).}
  \label{fig:sdg-query-example}
% \vspace{-3mm}
\end{figure*}

Fig.~\ref{fig:overview} shows an overview of the \toolname approach
for checking OAuth properties for Java or Javascript programs.
\toolname \ takes two inputs: (1) the source or byte code of an
application, and (2) a user-provided OAuth property $\query$ specifying the correct behavior using our query language.  Given these two inputs, \toolname \
performs signature matching by checking whether there exists an \emph{embedding} of the application with respect to property $\query$. 
% The callgraph automaton is translated from the initial callgraph of the application. To efficiently decide whether an application satisfies the precondition of an OAuth property, \toolname \ constructs the product of the query and callgraph automata.
% If the product automata is empty, we have \emph{refuted} the query, meaning that the callstack
% configuration specified by the user is not feasible.}

% On the other hand, if the product automata is non-empty, then \toolname proceeds to refine the product automaton
% and compute its corresponding system dependence graph (SDG). Here,
% the correct usage
% of the OAuth APIs are specified as queries, and
% provided to the signature matching tool, which matches the
% signatures against the system dependence graph.
% If a violation is found, a vulnerability is
% reported.
\subsection{Code Representation}~\label{sec:sdg}
Given a program, \toolname first generates its abstract representation using static analysis. In particular, 
we leverage \emph{System Dependence Graph (SDG)}, which summarizes both data- and control-dependencies among all the statements and predicates in the program. 

More formally, \texttt{SDG} for an application $A$ is a graph $(V, X, Y)$ where:

\begin{itemize}
\item $V$ is a set of vertices, where each $v \in V$  is a program statement of $A$.
\item $X$ encodes control-dependency edges. Specifically, $(v, v') \in X$ indicates that 
during execution, $v$ can directly affect whether $v'$ is executed.
Precisely, \texttt{SDG} creates three additional edges to handle function calls: (1) call edge, (2) parameter-in edge, and (3) parameter-out edge. Call edge connects the node at callsite (in caller) to the entry node of the called procedure (i.e., callee). Parameter-in edges connect the actual-in (caller) parameter nodes to the formal-in (callee) parameter nodes of the called procedure, and parameter-out edges connect the formal-out (callee) nodes to the actual-out (caller) nodes. 
\item $Y$ is a set of data-dependency edges. In particular, $(v, v', d) \in Y$ indicates that 
 statement $v$ and $v'$ are related by metadata $d$. Here, we use metadata $d$ to denote \emph{taint sources} that will be propagated by the data-flow analysis. Also, $X$ is data dependent on $Y$ if $Y$ is an assignment and
the value assigned in $Y$ can be referenced from $X$.
\end{itemize}

\vspace{0.05in}

\begin{example}
% Fig.~\ref{fig:sdg-query-example} (left) shows a simplified SDG constructed from the implementation (Fig.~\ref{fig:code-example-motivating}) of  \textit{node-oaut2-server}~\cite{lib-node-oauth2-server} library that shows an incorrect redirect URI validation at the authorization endpoint. Particularly, instead of matching the submitted URI with the pre-registered client's URI, the code example fetches the redirect URI from the pre-registered value whenever the submitted URI is missing. In the SDG, the nodes represent the statements of the program such as function calls, field access, etc. The edges represent the control and data-dependencies between the statements. For example, since the \textit{if condition} (line 7) depends on the value of \textit{redirect\_uri} field (line 4), the conditional node (\textit{==Null?}) has a data-dependency edge from the field access node (\textit{request.redirect\_uri}). On the other hand, control-dependency between the function call nodes (\textit{Get\_client()} and \textit{Auth\_req()}) implies the \textit{Get\_client()} is invoked after the \textit{Auth\_req()} is invoked.
Fig.~\ref{fig:sdg-query-example}(A) shows a simplified SDG constructed from the code example in Fig.~\ref{fig:code-example-motivating}. In the \texttt{SDG}, the nodes represent statements of the program such as function calls, field access, etc. The edges represent the control- and data-dependencies between the statements. For example, since the \textit{if condition} (line 7) depends on the value of \textit{redirect\_uri} field (line 4), the conditional node (\textit{==Null?}) has a data-dependency edge from the field access node (\textit{request.redirect\_uri}). On the other hand, control-dependency between the function call nodes, \textit{GetClient()} and \textit{AuthRequest()}, implies the \textit{GetClient()} is invoked after the \textit{AuthRequest()} is invoked.
\end{example}

\subsection{Facts and Inference Rules for SDG}~\label{sec:check}

Inspired by \oauthlint's formalization~\cite{DBLP:conf/kbse/RahatFT19}, \toolname converts the application's SDG into its corresponding facts and rules using Datalog. We first
give some preliminaries on Datalog program, and then describe the syntax and semantics of \toolname's built-in predicates.
% \redtext{RR2:Give proper credit to [55] regarding datalog predicates and remove overlapping text with the said work. }
% \paragraph{Datalog Preliminaries}
% A Datalog program  consists of a set of \emph{rules} and a set of \emph{facts}. 
% Facts simply declare predicates that evaluate to true. For example, 
% {\tt parent("Bill", "Mary")} states that Bill is a parent of Mary. Each  Datalog rule is  a Horn clause defining a predicate as a conjunction of other
% predicates. For example, the  rule:

% {\small
% \begin{verbatim}
% ancestor(x, y) :- parent(x, z), ancestor(z, y).
% \end{verbatim}
% }

% \noindent
% says that \verb+ancestor(x, y)+ is true if both \verb+parent(x, z)+ and
% \verb+ancestor(z, y)+ are true. In addition to variables, predicates can 
% also contain constants, 
% which are surrounded by double quotes, or
% ``don't cares",  denoted by underscores.

% Datalog predicates naturally represent relations. Specifically, if tuple $(x, y, z)$
% is in relation $A$, this means the  predicate $A(x, y, z)$ is
% true. In what follows, we write the type of a relation $R \subseteq X \times Y \times \ldots$ as $(s_1 : X, s_2 : Y, \ldots)$, where
% $s_1$, $s_2$, $\ldots$ are descriptive texts for the corresponding domains.
A Datalog~\cite{datalog-wiki} program is a set of \emph{facts} and \emph{rules} written in a declarative logic language. Facts correspond to predicates that evaluate to \emph{true}. For example, in SDG's context,
{\tt edge($Stmt_1$, $Stmt_2$)} implies that statement node $Stmt_1$ and $Stmt_2$ are connected by an edge. Each rule is a Horn clause~\cite{van1976semantics} defining a predicate as a conjunction of predicates. For example, the following program:

{\small
\begin{verbatim}
path(x, y) :- edge(x, y).
path(x, y) :- path(x, z), edge(z, y).
\end{verbatim}
}

\noindent
says that {\tt path(x, y)} is true if {\tt edge(x, y)} is true, or both {\tt path(x, z)} and
{\tt edge(z, y)} are true. In addition to variables, predicates can 
also contain constants (surrounded by double quotes), or
``don't cares",  denoted by `\_'.

\subsubsection{Base Facts in \toolname}~\label{subsec:base-facts}
Unlike \oauthlint whose analysis is flow-insensitive, many crucial properties in \oauthserver require a flow-sensitive analysis. Therefore, \toolname's facts take the form of $A(L,y, x_1, ... ,x_n)$, where $A$ is 
the instruction name,  $L$ is the instruction’s label, \texttt{y} is the variable storing
the instruction result (if any), and $x_1, ... ,x_n$ are variables given to
the instruction as arguments (if any). For example, the instruction
 $l_1: r_1 = 0$  is encoded as $\texttt{assign}(l_1, r_1, 0)$. Additionally, $\texttt{alloc}(l_1, y,x)$ means that variable $x$ may point to abstract location $y$ and $\texttt{alias(x,y)}$ denotes that variable $x$ and $y$ may point to the same abstract location. Furthermore, 
%  the store instruction $\texttt{store}(L,d, v)$ denotes that the value
%  of $v$ is stored in location $d$. 
 the branch instruction \texttt{branch}$(L_1,X, L_2,L_3)$ denotes that if $X$ is evaluated to \texttt{true}, then the next instruction will be $L_2$, otherwise $L3$. Using the base facts described above, \toolname computes the semantic facts of: (i) control-dependency predicates, which
capture instruction dependencies according to the application’s CFG,
and (ii) data-dependency predicates.

\subsubsection{SDG Predicates} 
\toolname provides built-in predicates to encode SDG generated from Sec.~\ref{sec:sdg}. 
% As shown in Fig.~\ref{fig:oauth-rule}, 
In particular, the $\mathtt{depOn}(y, x)$ predicate indicates that the value of variable $x$ has data-dependence on $y$. Similarly, the $\mathtt{followBy}$ predicate is inferred from the application’s CFG. Intuitively, $\mathtt{followBy}(L_1, L_2)$ holds for $L_1$ and $L_2$ if both are in the same
basic block and $L_2$ follows $L_1$, or there is a path from the
basic block of $L_1$ to the basic block of $L_2$.
The SDG predicates are computed using the following datalog rules:
    \[\begin{array}{rlll}
      \mathsf{depOn}(x, y) & :- & \mathsf{alloc}(\_,y, x)\\
      \mathsf{depOn}(x, y) & :- & \mathsf{assign}(\_,y, x)\\
      \mathsf{depOn}(x, z) & :- & \mathsf{assign}(\_,y, x), \mathsf{depOn}(y, z)\\
      \mathsf{depOn}(x, z) & :- & \mathsf{alias}(y,z), \mathsf{depOn}(x, y) \\
      \mathsf{followBy}(x, y) & :- & \mathsf{follow}(x,y) \\
      \mathsf{followBy}(x, z) & :- & \mathsf{followBy}(y,z), \mathsf{follow}(x, y) 
    \end{array}\]
Here, we use the $\mathtt{follow}(L_1, L_2)$ as the base case which holds if $L_2$ immediately follows $L_1$ in the CFG.

\subsubsection{OAuth Predicates}~\label{subsec:oauth-preds}
In addition to basic facts from the SDG, \toolname's query language also defines a list of predicates specific to the OAuth domain.
The predicate $\mathtt{OAuthTag}(L,T)$
defines that, the value at label $L$ is assigned tag $T$. Here, tag $T$ are associated with program statements that hold OAuth-specific resources such as \emph{redirect\_uri}, \emph{access\_token}, etc. For instance, the field access at line 8 in Fig.~\ref{fig:code-example-motivating} is denoted as $\mathtt{OAuthTag}(8, \mathtt{client\_redirect\_uri})$.
% As shown in Fig.~\ref{fig:oauth-rule},
% \texttt{isToken(x)} denotes that variable $x$ may point to an object that is an access token. 
Similar to~\cite{DBLP:conf/kbse/RahatFT19}, 
the $\mathtt{OAuthTag}$ predicate is computed through a standard data flow analysis in WALA. Specifically, our analysis first marks a set of APIs that could return OAuth resources as the sources. All variables assigned (either direct or transitive) by the sources will also point to the corresponding sources. In particular, 
since it is difficult
to precisely pinpoint strings that correspond to redirected URLs, we use both 
pattern matching (i.e., regular expressions) and domain-specific knowledge (i.e., API that may return a redirected URI or object values accessed by the keys defined in the specification) to over-approximate the domain of URI. Furthermore, $\mathtt{invoke}(L,m)$ predicate implies that method $m$ is invoked at a location with label $L$. Here, $m$ can be both OAuth-specific methods (e.g., $\mathtt{gen\_token}$) and programming language-specific methods (e.g., $\mathtt{matches}$). $\mathtt{sload}(L,q)$ predicate defines the load data operation for server's storage model $q$ (e.g., database). Similarly, $\mathtt{sstore}$/$\mathtt{sdelete}$ defines the store/delete operations. Finally, $\mathtt{error}(L,e)$ implies that an OAuth exception is thrown with an error message $e$. 
% Similarly, we also have other tags to denote \textit{access token}, \textit{code verifier}, and \textit{code challenge}, etc.

\begin{table*}[ht]
\centering
\footnotesize
\begin{tabular}{|p{0.45cm}|p{0.9cm}|p{6cm}|p{8cm}|}
\hline
\textbf{Prop.}
& \multicolumn{1}{c|}{\textbf{Grants}}
& \multicolumn{1}{c|}{\textbf{Description}}
& \multicolumn{1}{c|}{\textbf{\toolname query}}
\\ \hline
\textbf{P1}
& AuthCode, Implicit
& \oauthserver must validate the redirect URI parameter in the authorization request exactly matches with the client's registered URI before sending the redirected response to the URI. (RFC-6749)
& {\footnotesize
\begin{verbatim}
invoke(L1, auth_req),OAuthTag(L2, req_URI), OAuthTag(L3,client_URI), 
invoke(L4,redirect), error(L5, _), branch(L6,X,L4,L5),
followBy(L1,L2), followBy(L1,L3), depOn(L2,X), depOn(L3,X).
\end{verbatim}
}
\\ \hline
\textbf{P2}
& AuthCode, Implicit
& Redirect URI parameter in authorization request must be an \textit{absolute} URI. \oauthserver implementations utilize pattern matching APIs to perform this validation. (RFC-6749, RFC-6819).
& {\footnotesize
\begin{verbatim}
OAuthTag(L1, auth_req), OAuthTag(L2, req_URI), OAuthTag(L3,abs_URI), 
invoke(L4,matches), invoke(L5,redirect), error(L6,_),
branch(L7,X,L5,L6), followBy(L1,L2), depOn(L2,L4), depOn(L3,L4), 
depOn(L4,X).
\end{verbatim}
}
\\ \hline
\textbf{P3}
& AuthCode
& Authorization $\mathtt{code}$ must be single-use, meaning the $\mathtt{code}$ must be revoked once it is exchanged for token. If a revoked $\mathtt{code}$ is used again, \oauthserver must deny the request with an error message. (RFC-6749).
& {\footnotesize
\begin{verbatim}
invoke(L1, token_req), OAuthTag(L2, code), sload(L3, db),invoke(L4, 
gen_token), sdelete(L5, db), error(L6, _), branch(L7,X,L4,L6), 
followBy(L1,L2), depOn(L2,L3), depOn(L3,X), followBy(L4,L5), 
depOn(L2,L5).
\end{verbatim}
}
\\ \hline
\textbf{P4}
& AuthCode
& Authorization $\mathtt{code}$ must be bound to a certain client. Before issuing the token from the token request endpoint, \oauthserver must check the $\mathtt{code}$ is the same as the one issued to the client. (RFC-6749, OAuth Security Best Practices).
& {\footnotesize
\begin{verbatim}
invoke(L1, token_req), OAuthTag(L2, code), OAuthTag(L3,client_code), 
invoke(L4, gen_token),error(L5, _), branch(L6,X,L4,L5),
followBy(L1,L2),followBy(L1,L3), depOn(L2,X), depOn(L3,X).
\end{verbatim}
}
\\ \hline
\textbf{P5}
& AuthCode
& Authorization $\mathtt{code}$ must be bound to the client's redirect URI to where it was issued to. Before issuing the token, the server must check that the $\mathtt{code}$ is associated with the client's redirect URI. (RFC-6749).
& {\footnotesize
\begin{verbatim}
invoke(L1, token_req), OAuthTag(L2, code_URI), OAuthTag(L3,
client_URI), invoke(L4,gen_token), error(L5, _), branch(L6,X,L4,L5),
followBy(L1,L2), followBy(L1,L3), depOn(L2,X), depOn(L3,X).
\end{verbatim}
}
\\ \hline
\textbf{P6}
& AuthCode
& \oauthserver must store the PKCE parameters (i.e., code challenge and code challenge method) at the authorization endpoint to be validated at token endpoint (RFC-7636, OAuth Security Best Practices).
& {\footnotesize
\begin{verbatim}
invoke(L1, auth_req), OAuthTag(L2, code_challenge), OAuthTag(L3,
code_challenge_method), sstore(L4,db), followBy(L1,L2),
followBy(L1,L3), depOn(L2,L4), depOn(L3,L4).
\end{verbatim}
}
\\ \hline
\textbf{P7}
& AuthCode
& \oauthserver must verify the authenticity of PKCE parameters at token endpoint. The transformed (i.e., sha256) value of $\mathtt{code\_verifier}$ parameter must be same as the $\mathtt{code\_challenge}$ value (RFC-7636, OAuth Security Best Practices).
& {\footnotesize
\begin{verbatim}
invoke(L1, token_req),OAuthTag(L2, code_challenge), OAuthTag(L3, 
code_verifier), invoke(L4, sha256), invoke(L5, gen_token),
error(L6, _), branch(L7,X,L5,L6), followBy(L1,L2), followBy(L1,L3), 
depOn(L3,L4), depOn(L4,X), depOn(L2,X).
\end{verbatim}
}
\\ \hline
\textbf{P8}
& AuthCode, Implicit
& \oauthserver must provide CSRF protection by handling the $\mathtt{state}$ parameter in authorization request. Value of the $\mathtt{state}$ parameter must be added to the redirected response of authorization request. (RFC-6749, OAuth Security Best Practices).
& {\footnotesize
\begin{verbatim}
invoke(L1, auth_req), OAuthTag(L2, state), invoke(L3, redirect), 
error(L4,_), branch(L5,X,L3,L4), followBy(L1,L2), depOn(L2,X), 
depOn(L2,L3).
\end{verbatim}
}
\\ \hline
\textbf{P9}
& AuthCode, Implicit
& Access tokens issued by the \oauthserver should be constrained to a certain client. \textit{m-TLS} is a standardized and widely used client-constrained mechanism, in which the server first obtains the client’s certificate from TLS stack, decodes and hashes the certificate and finally, associates it with the access token. (RFC-8705, OAuth Security Best Practices).
& {\footnotesize
\begin{verbatim}
OAuthTag(L1, access_token), OAuthTag(L2, client_cert),
invoke(L3, b64_decode), invoke(L4, b64_encode), invoke(L5, sha256), 
invoke(L6, add_cert), depOn(L2,L3),depOn(L3,L5),depOn(L5,L4),
depOn(L1,L6),depOn(L4,L6).
\end{verbatim}
}
\\ \hline
\textbf{P10}
& AuthCode, Implicit
& \oauthserver should not store access tokens as clear-text and should store access token hashes only (RFC-6819).
& {\footnotesize
\begin{verbatim}
OAuthTag(L1, access_token), invoke(L2, sha256), sstore(L3,db), 
depOn(L1,L2), depOn(L2,L3).
\end{verbatim}
}
\\ \hline
\end{tabular}

\caption{We define ten security properties to detect the most commonly observed logical flaws on OAuth servers based on the standard OAuth specifications (RFC-6749~\cite{rfc6749}, RFC-6819~\cite{rfc6819}, RFC-7636~\cite{rfc7636}) and OAuth security current best practices~\cite{oauth-best-practice}. We express the \toolname queries using the predicates defined in Sec.~\ref{sec:check}.} 
% For \oauthserver, five (P4, P5, P6, P7, and P9) of these security properties are studied by us for the first time.}
\label{tab:list-of-properties}
\vspace{-6mm}
\end{table*}

\subsection{Query Language}
\label{sec:query}
In this section, we show how to express OAuth properties over
semantics facts of an application. We begin by defining the query language for
expressing security patterns. 
% We continue by presenting a set of relevant security properties, and for each, we show
% compliance and violation patterns, which imply the property and,
% respectively, its negation. 
This construction enables us to determine whether an application complies with a given security
property. Specifically, for each OAuth property, the user defines a
unique predicate that serves as the signature for the property. The user may also define additional helper predicates used by the signature. 

\textbf{Syntax.} The syntax of the query language is given by the
following Backus–Naur form (BNF):
	\[\begin{array}{rlll}
% 		v_i & ::= & \mathsf{arg} ~|~ \mathsf{reg} ~|~ \mathsf{mem} ~|~ ... \\
		\pred & ::= & \mathsf{depOn}(v_1, v_2) ~|~ \mathsf{followBy}(v_1, v_2) ~|~ \mathsf{branch}(...) \\
		&|& 
		\mathsf{OauthTag(x,y)} ~|~ \mathsf{invoke(l,m)} ~|~ \mathsf{sload(l,q)} ~|~ \mathsf{...} \\
        &|& \neg \pred ~|~ \pred \land \pred  ~|~ \pred \lor \pred
    \end{array}\]
We now formally define our \emph{property signatures} and state what it means for an app to \emph{match} a property.
Intuitively, a signature for a property $\mathcal{F}$ is an \texttt{SDG} $(V_0, X_0,Y_0)$ %(where $(X_0=V_0,E_0)$)
that captures  semantic properties.
Ideally, $G_0 = (V_0, X_0,Y_0)$ would satisfy the following: $G_0$ occurs as a subgraph (defined below) of the \texttt{SDG} $G_S = (V_S, X_S,Y_S)$ of an OAuth application.

By ``occurs as a subgraph'', we mean  there exists an embedding
%\begin{align*}
$F_S:V_0\to V_S$
%\end{align*}
such that the following properties hold:
\begin{itemize}
\item {\bf One-to-one.} For every $v,v'\in V_0$ where $v\not=v'$, $F_S$ cannot map both $v$ and $v'$ to the same vertex, i.e.,
$
F_S(v)\not=F_S(v').
$
\item {\bf Type preserving.} For every $v\in V_0$, $F_S$ must map $v$ to a vertex of the same type, i.e., $
T(v)=T(F_S(v))$
\item {\bf Edge preserving.} For every $v,v'\in V_0$, $F_S$ must map an edge $(v,v')\in X_0$ to an edge in $X_S$:
\begin{align*}
(v,v')\in X_0\Rightarrow (F_S(v),F_S(v'))\in X_S.
\end{align*}
\end{itemize}

Given property $G_0 = (V_0, X_0, Y_0)$ and  app $S$ with \texttt{SDG} $G_S = (V_S, X_S, Y_S)$,  we say that $G_0$ \emph{exactly matches} (or simply \emph{matches}) $S$ if $G_0$ occurs as a subgraph of $G_S$. In other words, given a signature $(V_0,X_0,Y_0)$ and a sample $S$ with \texttt{SDG} $(V_S,X_S,Y_S)$, we can check whether $(V_0,X_0,Y_0)$
matches $S$. If so, we have determined that $S\in\mathcal{F}$; otherwise, $S\not\in\mathcal{F}$.

% \redtext{RR3:Bring to the main text a summarized explanation of the formalized OAuth security properties presented in Table 2.}

\subsubsection{Expressing OAuth properties.}~\label{subsec:express-properties}
Table~\ref{tab:list-of-properties} shows the list of properties that we express using our query language. We obtain these properties from the documentation of standard OAuth specifications~\cite{rfc6749, rfc6819,rfc7636} and security best practices~\cite{oauth-best-practice}. In what follows, we briefly discuss the properties and their formal representation. 
% Interested readers can find more details of these properties, including their security implications in Appendix~\ref{apx:properties} as well as corresponding attack examples in Appendix~\ref{apx:attacks}.

%In what follows, we briefly discuss and formalize \emph{Redirect URI} property discussed in Section~\ref{sec:background}. More details about other properties in Table~\ref{tab:list-of-properties} and their security implications are described in the %Appendix~\ref{apx:properties}

% To explain the properties, we use code examples adapted from the open-source implementation of OAuth server. In the following, we describe how we express these properties using our query language:

OAuth requires that the redirect URI submitted during the authorization request match the client's registered URI (P1). Specifically, we tag these OAuth parameters by $\mathtt{OAuthTag(\_,req\_URI)}$ and $\mathtt{OAuthTag(\_,client\_URI)}$, respectively. In addition, since this URI is used to transfer sensitive information like authorization code, the specification also requires the URI to be an \emph{absolute URI} (P2), preventing attackers from utilizing the open redirector of the user agent to intercept the sensitive information. Server programs commonly use pattern matching APIs (e.g., `\texttt{matches}') to match the redirect URI value.

To prevent code replay attacks during token request, OAuth also requires the authorization code to be single-use (P3), meaning the code can be used only once (to generate token) by the client. Therefore, once the code is used, \oauthserver removes it from the storage (e.g., database) and rejects requests (with error message) when the code is not found in storage. We use $\mathtt{OAuthTag(\_,code)}$ to tag authorization code and \texttt{invoke(\_,gen\_token)} to track the OAuth-specific API call for generating token.

OAuth security best practices require \oauthserver supporting PKCE~\cite{rfc7636} to protect public clients (e.g., native desktop apps). To securely implement PKCE, servers first receive and store the PKCE parameters (e.g., \texttt{code\_challenge}) at authorization endpoint (P6), and transforms (by `\texttt{sha256}') the \texttt{code\_verifier} parameter to match with the \texttt{code\_challenge} (P7) at the token endpoint. 

% The \oauthserver should also support using the $\mathtt{state}$ parameter (P8), a string value that helps to maintain the state between the client's request and response at the authorization endpoint. 

OAuth also requires access token to be constrained to a certain client to prevent token injection attacks (P9). Mutual-TLS~\cite{rfc8705} (also noted as m-TLS) is a standardized and widely used client-constrained mechanism that allows the clients to demonstrate the proof of possession when using the access token. In m-TLS, the server first obtains the client's certificate from TLS stack and associates it with the token before sending to the clients. The certificate is decoded and encoded using standard Base64 decode-encode APIs. In our formalization, we use $\mathtt{OAuthTag(\_,client\_cert)}$ to tag statements holding client's certificate, and $\mathtt{invoke(L4, b64\_encode)}$, $\mathtt{invoke(L4, b64\_encode)}$, etc. to track the corresponding domain-specific APIs.
Once a query expressing a property from the OAuth specification is submitted, \toolname checks the it against the SDG representation of the input program.
\begin{example}
% \noindent
% \textbf{Signature matching}
Fig.~\ref{fig:sdg-query-example}(B) shows a graphical representation of the query that \emph{over-approximates} the redirect URI property (P1) in Table~\ref{tab:list-of-properties}. It represents a signature where a redirect URI received during an authorization request (\textit{request.redirect\_uri}) is matched with the redirect URI field of a client instance (\textit{client.redirect\_uri}) before making a \textit{redirect} function call, which must be used for a successful response to the authorization request. Therefore, this signature can be used to check that a successful redirection occurs only after the check for redirect URIs is performed. Otherwise, an error response is generated. On the other hand, the SDG (Fig.~\ref{fig:sdg-query-example}(A)) of the program for processing authorization request  retrieves the `\textit{redirect\_uri}' from the `\textit{request}' instance, but the `\textit{redirect\_uri}' is matched only against the \textit{NULL} value before sending the redirection using `\textit{sendRedirect}' function call. However, OAuth specification requires that the server must check whether the `\textit{redirect\_uri}' of the authorization request is the same as the `\textit{redirect\_uri}' registered to the client, which is not done in this program. Therefore, \toolname reports a violation.
\end{example}

\section{Implementation}~\label{sec:impl}
This section discusses the design and implementation of \toolname, as
well as a few key optimizations.

\subsection{Query-driven Exploration}
\label{sub:query-driven}
% \redtext{RR4: Provide more technical details on the design of Cerberus, in particular, the query-driven slicing technique which seems to be instrumental for improving scalability; to help the explanation, replace the "foo-bar" CGA in Fig 6 and 7 with the CGA of a concrete OAuth program, if possible the running example in Fig 3.}
\toolname leverages System Dependence Graph (SDG) that captures program's control- and data-flow dependencies. 
However, there is a steep trade-off between the precision of an SDG and the cost of constructing it. For example, SDGs 
constructed using a context-insensitive pointer analysis tend to grossly overapproximate the targets of virtual method calls, leading to unacceptable false alarms. On the other hand,
more precise SDGs obtained using context-sensitive
pointer analysis can take hours to construct.  Currently, analyses that rely on system dependence graph information must implement their own ad hoc analysis~\cite{he2015vetting, egele2013empirical} to answer application-specific queries. \yu{ To mitigate this challenge, we introduce a query-driven approach whose key insight is to only reason about small program fragments relevant to the OAuth property.}

Since the original system dependence graph is obtained by stitching all control flow graphs from the methods, our key intuition is to keep track of the methods relevant to the OAuth property. In particular, we define OAuth request endpoint $Q$, \yu{which is the precondition of the actual OAuth property} specifying the entry and exit points of the relevant code snippet. After that, we leverage the request endpoints to pinpoint a sub-callgraph between the request endpoints. Since the sub-callgraph is typically small, its corresponding system dependence graph will also be small.

% Figure~\ref{fig:example-cga-qa} illustrates our approach to compute the sub-callgraph given the OAuth endpoints. While the standard OAuth endpoints (e.g., token endpoint) are pre-defined in \toolname, we allow developers to provide additional endpoints as input using regular expressions (shortened as \textit{regex}). 
Given an endpoint regex as input, we implement a lightweight \emph{program slicing} using automata theory~\cite{dragon-book}. Specifically, \toolname \ first constructs the so-called
\emph{query automaton (QA)}~\cite{explorer} for the request endpoints and the \emph{callgraph automaton (CGA)}.
Here, the query automaton is simply an NFA-representation of the regular
expression specified by the user.  Since the problem of converting regular
expressions to finite state machines is well-studied~\cite{berry1986regular}, we do not explain
the QA construction in detail here.

We now explain the syntax and semantics of its query language for specifying request endpoints.  For a given program $P$, \toolname \ accepts
specifications written in the following query language:
\vspace{-1.4mm}
\[
\begin{array}{lll}
\emph{Query} \ Q & := &  f \in {\rm methods}(P) \\ & &  | \ \ .  \ \ |  \ \ Q_1 \concat Q_2 \ \   \\
& & | \  Q_1 \disjunct Q_2 \ \ | \  \ Q^* \ \ | \ \ Q^+  \ \ | \ \ (Q) 
\end{array}
\]
% \vspace{-1mm}
The building blocks of queries are method names in program $P$, denoted by
${\rm methods}(P)$. The dot character (``$.$'') matches any method name,
and the $\concat$ operator indicates a call from one method to another.
% The exclamation mark operator ($``!"$) is used to negate queries and 
The
``$+$" operator is used for taking the disjunction of two queries. As usual,
the ``$*$'' operator stands for Kleene closure, and $Q^+$ is syntactic
sugar for $Q \rightarrow Q^{*}$.

The callgraph automaton CGA for a given application $P$ with respect to a callgraph $C$ is a finite state machine where states include all methods of $P$ and transition functions correspond to call edges (i.e., function calls.). After constructing the query and callgraph automata, the next step is to compute the \emph{intersection} of those two using the JSA~\cite{jsa} tool. The output is a \emph{product automata} that encode a relevant program slice with respect to the query. 

\begin{figure}[!t]
\centering
  \includegraphics[scale=0.45]{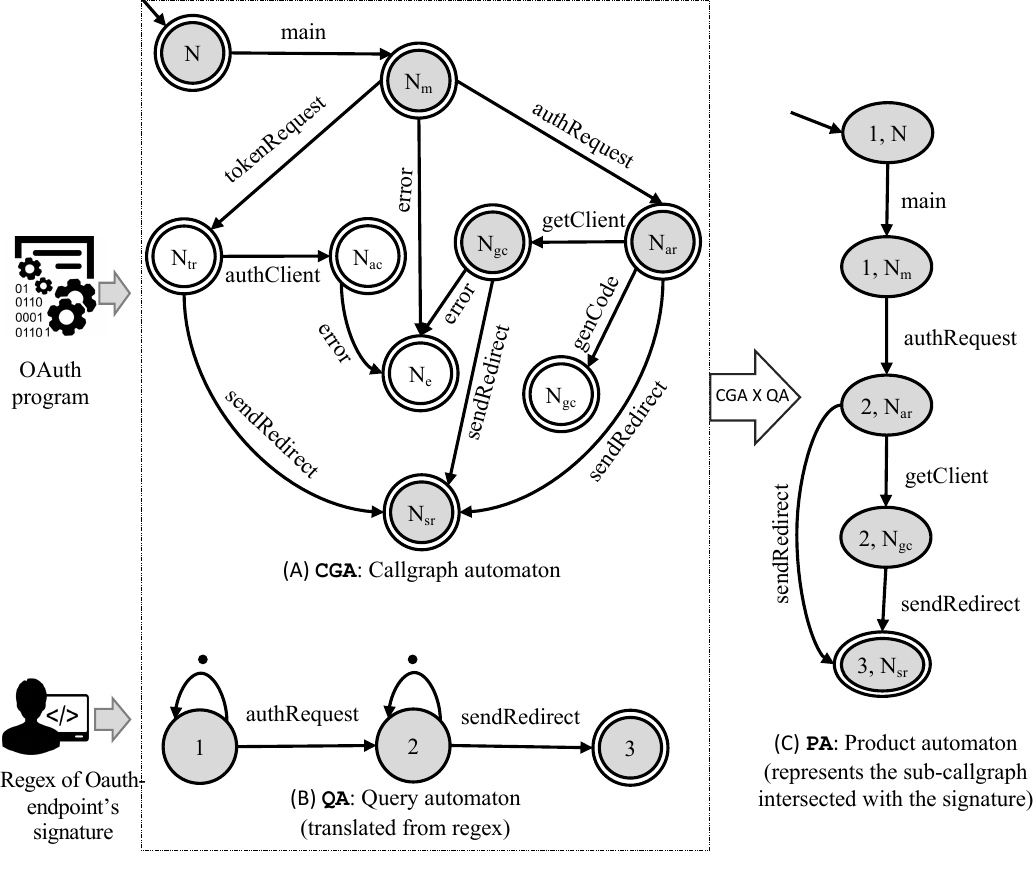}
  \caption{An illustration of our query-based slicing technique where (A) is a partial callgraph automaton (CGA) constructed for the OAuth program in Fig.~\ref{fig:code-example-motivating}, (B) is an example of a query automaton (QA) that we use to represent OAuth endpoint's signature, and (C) is the partial product automaton constructed by \toolname.}
  \label{fig:example-cga-qa}
\vspace{-6mm}
\end{figure}
% \begin{figure}[H]
% \centering
%   \includegraphics[scale=0.45]{figures/product-automaton.pdf}
%   \caption{\tamjid{Example  of partial product automaton constructed from CGA and QA in Fig.~\ref{fig:example-cga-qa} that is used to determine the sub-callgraph.}}
%   \label{fig:example-product-automaton}
% \end{figure}
\begin{example}
Fig.~\ref{fig:example-cga-qa} (A) shows a partial call graph Automaton (CGA) constructed from the OAuth program in Fig.~\ref{fig:code-example-motivating}. Since the original program may lead to a large call graph that is difficult for a whole-program analysis, the user can optionally obtain a relevant program slice by specifying a query ``$(.^{*} \rightarrow \mathtt{authRequest} \rightarrow .^{*} \rightarrow \mathtt{sendRedirect})$'', which is a regular expression representing the entry and exit point of OAuth endpoints. Here, our goal is to extract the intersection between the callgraph and the user-specified query. Fig.~\ref{fig:example-cga-qa} (B) shows a query automaton (QA) that is obtained from the previous regular expression through a standard algorithm based on induction~\cite{dragon-book}. Having constructed the query and callgraph automata, the next step is to determine whether the intersection of the two is empty. Fig.~\ref{fig:example-cga-qa} (C) shows a partial product automaton (PA) constructed by the CGA (A) and QA (B), i.e., $L(PA) = L(CGA) \cap L(QA)$ where $L(PA)$ represents the langauge accepted by product automaton $PA$, which is again can be computed by a standard algorithm from ~\cite{dragon-book}. Here, every state that would appear in the product automaton must also appear in both the call graph automaton and query automaton. Therefore, the resulting PA represents a \emph{program slice} that is relevant to the query.
% For example, an authorization endpoint that begins with an authorization request and ends with a redirection response can be represented using the query $(.^{*} \rightarrow X:auth\_req \rightarrow .^{*} \rightarrow X:redirect)$. 
% To compute the sub-callgraph relevant to a given query, \toolname constructs the product of the CGA and QA. As shown in Figure~\ref{fig:example-product-automaton}, partial product automaton that leads to an accepting state allows \toolname to compute the sub-callgraph relevant to the OAuth properties.
\end{example}
We note that, compared to a naive whole-program analysis, our query-driven exploration does not miss a property violation as long as all the necessary endpoints are provided. Since the endpoints' signature defining the entry and exit node is the precondition for the each property, \toolname will not prune any paths that might violate the corresponding property.

\subsection{Hybrid Analysis for Dynamic Features}
\label{sec:hybrid}
The system dependence graph may not capture the full semantics of some OAuth properties. For instance, one CVE entry (\textit{CVE-2020-26938}), identified by \toolname, is caused by checking if the redirect URI contains an absolute URI using an incorrect URI pattern \textit{"[a-zA-Z][a-zA-Z0-9+.-]+:"}. This may lead to incorrect results since our current abstraction does not reason about the semantics of regular expressions. Also, since our tool is based on static analysis, it may not determine whether a branch condition evaluates to true or not. However, a fully dynamic analysis will be prohibitive since we have to deal with libraries with a large codebase.

To mitigate the above-mentioned challenge, \toolname incorporates a hybrid approach: in particular, given an application that may potentially contain dynamic features or regular expressions that go beyond the scope of our current static analysis, we first make the most conservative assumption by assigning relevant predicates to \texttt{false}. For instance, the \texttt{branch} predicate will evaluate to false if its condition contains regular expressions, which will fail the signature matching procedure and raise a potential \emph{false alarm}. After that, we perform a light-weight \emph{delta testing} as follows: for each predicate that is assigned to \texttt{false} due to our conservative assumption, we dynamically exercise the relevant code to recover the missing facts. For example, for the case with regular expressions, given a set of input strings, we will test whether the actual regular expression's output is the same as the ones generated by the correct regular expression. If so, we turn its corresponding predicate to \texttt{true} and rerun \toolname. We iterate this process until \toolname confirms a violation or all false alarms are eliminated. Our current implementation can handle
cases of dynamic features (e.g., reflective calls in Java) where it takes arguments with string constants. In such cases, we leverage the current data-flow analysis to keep track of the strings that may be used as class or function names.

\subsection{The \toolname Tool}~\label{subsec:tool}
We implemented our core static analysis
on top of the WALA framework~\cite{wala}, which provides compilation and analysis infrastructure for both Java and Javascript. \toolname's implementation consists of approximately 9,860 lines of Java code. We use an Andersen-style pointer analysis~\cite{andersen1994program} and the CHA (Class Hierarchy Analysis) callgraph algorithm provided by WALA. An OAuth-endpoint expressed in regular expression, is converted into its query automata using the JSA library~\cite{jsa}.

% However, Javascript poses some additional challenges for pointer analysis due to its flexible object model. For example,
% \begin{itemize}[leftmargin=*]
%     \item Unlike like Java, Javascript does not have any built-in concept of object instance method. Instead, Javascript functions are first-class values, and methods are functions stored in object properties (corresponding to instance fields in Java).
%     \item Object properties in Javascript can be created dynamically and property values can be functions or any other objects. Therefore, the set of properties an object may hold is not evident from the code, unlike in a static language like Java.
% \end{itemize}
\noindent
\textbf{From source code to Datalog programs.} 
Given a service provider program $P$ written in Java or Javascript, \toolname first leverages the WALA framework to generate its corresponding system dependence graph (SDG) from bytecode (for Java) or scripts (for Javascript). As shown in Figure~\ref{fig:sdg-query-example}, each node of the SDG corresponds to a statement using WALA's intermediate representation in Static Single Assignment (SSA) form. Each statement will be translated into its Datalog fact for allocation, assignment, function call, etc. Second, each edge is translated into it corresponding predicate in Datalog. For instance, the control- and data-dependence edges are translated into their corresponding $\mathtt{followBy}$ and  $\mathtt{depOn}$ predicates discussed in Section~\ref{sec:check}, respectively.
% To capture the semantics of dynamic object properties Javascript programs, we utilize WALA's  technique which tracks the correlations between dynamic property declaration and access that use the same property name. \toolname incorporates WALA's callgraph and pointer analysis as well as 
% its intermediate representation to construct the SDG for the input programs. 
% \revA{2} Since Java and Javascript programs are both converted into the abstract intermediate representation in Static Single Assignment (SSA) form, they both can be translated into our Datalog facts defined in Sec.~\ref{sec:check}. 
 Finally, \toolname leverages the Souffl\'e~\cite{souffle} Datalog solver for checking conformance between the SDG and the OAuth properties. 

% \paragraph{Callgraph construction}
% Our analysis begins with generating the callgraph on-the-fly from the program represented as byte code (Java) or source code (Javascript). To generate the callgraph for Javascript libraries, the source code is first represented in Abstract Syntax Trees (AST). We use Rhino to parse the Javascript source code and build ASTs. The generated ASTs are then used to build an Intermediate Representation (IR) of the code Static Single Assignment (SSA) form. For Java programs, IR is built upon the byte code. A Control Flow Graph (CFG) is then built on top of the IR and a Class Hierarchy Analysis (CHA) is also performed. Then a context and field-sensitive pointer analysis is performed to compute the \textit{points-to} mapping for the variables in the program that are alias of each other. A callgraph is constructed simultaneously during the pointer analysis. 
\noindent
\textbf{Resolving \textit{Node.js} modules.} For Javascript, the callgraph is constructed directly from the source code (i.e., scripts). Therefore, function calls from an included module from \textit{Node.js} framework is not automatically resolved as the required source files are not known to the analysis. For example, to implement a HTTP server in \textit{Node.js} framework, developers may call \textit{http.createServer()} from the HTTP module by using \textit{require(`http')}. Therefore, to resolve the call \textit{createServer()}, we first need to identify the required source file to be included in the analysis. We use the pointer analysis to identify the strings that can flow to a \textit{require} call. Then the corresponding file is loaded in the analysis, and the target method is included in the callgraph.
\section{Evaluation}~\label{sec:eval}
To determine the effectiveness of \toolname, we evaluate  it on popular OAuth-server libraries to answer the following research questions:
\begin{itemize}
    \item \textbf{RQ1}: 
    Can \toolname identify real-world vulnerabilities?
    \item \textbf{RQ2:} 
    Is our query-driven approach efficient and effective?
    % \item \textbf{RQ3}: Can we find real-world vulnerabilities in the popular and widely used OAuth-server implementation?
\end{itemize}

% Our evaluation shows that \toolname can effectively discover the OAuth vulnerabilities and substantially improve the query answering time compared to the eager analysis approach.
\subsection{Experimental Setup}\label{subsec:setup}
We conduct all experiments on a Quad-Core Intel Core i5 computer and 16GB of memory running on the macOS 12.1 operating system. In what follows, we elaborate on the details of the setup.

\subsubsection{Dataset}\label{subsec:dataset}

We consider high-profile OAuth-server libraries written in Java or Javascript. With extra engineering effort, our techniques can technically be applied to applications in other languages as well. To answer the research questions, we consolidate two datasets of open-source \oauthserver libraries that implement the standard OAuth specification~\cite{rfc6749}.
\begin{itemize}[leftmargin=*]
\item \datasetUnknown: This dataset contains 10 popular open-source libraries of OAuth server (i.e., service provider) where it is \emph{unknown} whether the libraries satisfy the security properties. Table~\ref{tab:library-stats} shows the key statistics for the selected libraries. To select the libraries, we first consider their popularity among the web developers. For example, \textit{Node Oauth2 Server}~\cite{lib-node-oauth2-server} library has more than 200k downloads each month. Similarly, \textit{Oauth2orize}~\cite{lib-oauth2orize} and \textit{Node oidc provider}~\cite{lib-node-oidc-provider} have approximately 179k and 65k downloads each month. Additionally, we also consider the number of dependant repositories (i.e., repositories that use the APIs of the library) as an indicator of the popularity of our selected libraries. All of the libraries support the widely used authorization code grant~\cite{rfc6749}, and three also support the implicit grant. Four of these libraries are implemented in Javascript, and six are implemented in Java. Some libraries (e.g., \textit{OxAuth}) are considerably larger than others as they also support additional endpoints (e.g., token introspection) and platform-specific custom grants for their authorization server. 

\item \datasetKnown: This dataset contains 12 OAuth libraries with 49 OAuth-specific logical flaws that are confirmed by Github issues or online forums. Among the 49 logical flaws, 13 were caused for improper handling of authorization code, 12 for mishandling access token, 7 for redirect URI validation, 8 for missing or incorrect PKCE validation, 3 for state parameter, and 6 for other issues such as incorrect client validation.  
% In total, the dataset contains 49 known vulnerabilities all of which are confirmed by the developers in publicly available platforms (e.g., github issues, stackoverflow, etc.). 
While the selection criteria for these libraries are similar to \datasetUnknown, we exclude libraries whose ground truths are unknown. Six of these libraries are implemented in Javascript, and six are implemented in Java. All libraries follow the standard OAuth specifications and support all the commonly used grants.
\end{itemize}

\begin{table}[t]
\centering
\footnotesize
\begin{tabular}{|l|c|c|c|c|}
\hline
\textbf{OAuth-server libraries}
& \textbf{Ver.}
& \textbf{Language}
% & \textbf{\#downloads (npm)} 
% & \textbf{\#stars (github)} 
& \textbf{Grants} 
& \textbf{\#LOC} 
\\ \hline

1) Node oauth2 server~\cite{lib-node-oauth2-server}
& 3.1.1
& JS
% & 187k/month
% & 3,180
& AC
& 3,936\\ 
\hline

2) Oauth2orize~\cite{lib-oauth2orize}
& 1.11.0
& JS
% & 179k/month
% & 3,119
& AC, I
& 3,483\\  
\hline

3) Node oidc provider~\cite{lib-node-oidc-provider}
& 6.29.5
& JS
% & 65k/month
% & 1,291
& AC, I
& 15,493\\ 
\hline

% node-oauth20-provider
% &
% &
% &
% &
% &\\ 
% \hline

4) Oauth2 server~\cite{lib-oauth2-server-node}
& 1.0
& JS
% & N/A
% & 105
& AC
& 2,946\\  
\hline

5) Spring auth server~\cite{lib-spring-authorization-server}
& 1.0
& Java
% & N/A
% & 2,978
& AC
& 28,374\\ 
\hline
6) Clouway server~\cite{lib-clouway-oauth2-server}
& 1.0.6
& Java
% & N/A
% & 37
& AC
& 8,462\\ 
\hline
7) Jobmission server~\cite{lib-job-oauth2-server}
& 1.0
& Java
% & N/A
% & 321
& AC
& 6,480\\ 
\hline

8) Apifest~\cite{lib-apifest-oauth20}
& 0.3.1
& Java
% & N/A
% & 67
& AC
& 14,371\\ 
\hline

9) Yoichiro server~\cite{lib-oauth2-server}
& 1.0
& Java
% & NA
% & 96
& AC
& 6,764\\ 
\hline

10) OxAuth~\cite{lib-oxAuth}
& 3.0.2
& Java
% & NA
% & 276
& AC, I
& 62,857\\  
\hline

\end{tabular}

\caption{Statistics for the open-source OAuth server libraries in \datasetUnknown. Here, AC and I indicate the OAuth's authorization code grant and implicit grant, respectively.}
\label{tab:library-stats}
\vspace{-6mm}
\end{table}

\subsubsection{State-of-the-arts.}~\label{subsec:sota} Existing tools focus on OAuth properties by analyzing the flows observed from client-side applications while treating the server-side as blackbox. Therefore, there is no prior work that directly performs whitebox analysis on \oauthserver programs. \oauthlint~\cite{DBLP:conf/kbse/RahatFT19} is the closest to our work as they also statically analyze source code of client applications to check OAuth properties. Unlike \toolname, \oauthlint only supports OAuth properties expressed via data-flow predicates. Furthermore, \oauthlint performs whole-program analysis over the client application. \skvetter~\cite{yang2018vetting} leverages symbolic execution to check security properties in OAuth SDKs, which provides APIs for implementing client applications. Similar to \oauthlint, \skvetter also consider server-side implementation as blackbox. In fact, \skvetter implements it's own \emph{model} of the server to analyze the client-side flows for all of the SDKs. Its properties are defined in terms of the request-response behavior observed from client apps. Therefore, we cannot compare against \skvetter because of unsupported properties (e.g., checking expressions, API calls, etc.) and unsupported languages (e.g., \skvetter only supports Python). 

% \redtext{RR7: Explicitly explain what human efforts are required to use Cerberus.}
\subsubsection{User inputs}~\label{subsec:userinput}
% Using \toolname to check the properties in OAuth applications is straightforward. 
% To begin the analysis, \toolname first requires an analyst to specify the entry points (e.g., \textit{main} function) of the application. Then the analyst can submit the queries for the OAuth properties they want to check.
\toolname takes as inputs a target program $P$ and a property $Q$. For each benchmark, we check it against all properties defined in Table~\ref{tab:list-of-properties}.
% To minimize the human effort, we provide a set of predicates that can be re-used to build queries for most of the OAuth properties. Since all the applications follow a standard protocol, these queries need to be specified only once and can be re-used across multiple applications. 
% Properties can be expressed using our query language defined in Sec.~\ref{sec:prob}. 
To quantify the manual effort of writing the queries before running the evaluation, we recruited 18 independent students from a graduate-level security class -- who were provided with the OAuth properties (i.e., from specification written in English) and documentation for our query language. We provide five randomly selected properties to each participant and measure the time they spent expressing the properties in our query language. We found participants spent on average 10.6 minutes to specify the given properties using our query language. 

To speed up the analysis, \toolname provides built-in OAuth endpoints (as defined in the specification) to leverage the query-driven slicing method (Sec.~\ref{sub:query-driven}). Meanwhile, we allow users to provide (as query) their own custom endpoints in regex format to check any additional endpoints that is not covered by the standard specification. Thanks to the library documentation, these endpoints' format are clearly specified for most libraries and can be easily defined using regex for our tool. This flexibility also allows the analysts to use \toolname to efficiently check security properties in extended or custom endpoints (e.g., token introspection) as well as new OAuth extensions (e.g., DPoP token~\cite{dpop-ietf}).
\subsection{Evaluation Results}\label{subsec:exp-results}

% [Note: to save space, the details of results bellow can be shorten, because we already present these results in Table 4.]
\subsubsection{Discovered Vulnerabilities} \label{subsec:discovered-vulnerabilities}
To answer \textbf{RQ1}, we use \toolname to identify logical flaws from both known and unknown datasets.

\noindent
\textbf{Detection of known vulnerabilities.} We first evaluate \toolname using \datasetKnown which consists of 12 popular OAuth libraries with 49 known flaws. 
% To identify the logical flaws using \toolname, we first encode the corresponding properties using our query language according to the reports from the public platforms. 
As shown in Table~\ref{tab:eval-ground-truth}, using the properties from Table~\ref{tab:list-of-properties}, \toolname identifies 43 out of 49 logical flaws.

\subsubsection{Unknown vulnerabilities} We next evaluate \toolname using  \datasetUnknown with ten popular open-source libraries of OAuth service providers.

% Commonly used static analysis tools, including WALA involve approximations, which may result in false-positive and false-negative cases in the analysis result. 
% To verify the results reported by \toolname for \datasetUnknown,

\begin{table}[h]
\centering
\footnotesize
\begin{tabular}{|>{\raggedright\arraybackslash}p{2.7cm}|>{\centering\arraybackslash}p{1.4cm}|>{\centering\arraybackslash}p{2cm}|>{\centering\arraybackslash}p{0.6cm}|}
\hline
\multirow{2}{*}{\textbf{OAuth-server libraries}} & 
\textbf{\#Known logical flaws}  & 
\textbf{\#Flaws identified by \toolname} & 
\multirow{2}{*}{\textbf{\#FN}} \\ \hline
1) OxAuth\cite{lib-oxAuth} & 9 & 8 & 1 \\\hline
2) Mitre Server\cite{lib-mitre} & 7 & 6 & 1 \\\hline
3) Spring Auth Server\cite{lib-spring-authorization-server} & 6 & 6 & 0 \\\hline
4) Node OAuth2 Server\cite{lib-node-oauth2-server} & 6 & 4 & 2 \\\hline
5) Loopback OAuth\cite{lib-loopback} & 4 & 4 & 0 \\\hline
6) OAuth Provider\cite{lib-node-oauth-provider} & 4 & 3 & 1 \\\hline
7) Egg OAuth2 Server\cite{lib-egg-oauth} & 4 & 4 & 0 \\\hline
8) ApiFest OAuth\cite{lib-apifest-oauth20} & 1 & 1 & 0 \\\hline
9) Clouway Server\cite{lib-clouway-oauth2-server} & 3 & 3 & 0 \\\hline
10) OAuth2orize\cite{lib-oauth2orize} & 3 & 2 & 1 \\\hline
11) Java OAuth Server\cite{lib-java-oauth-server} & 1 & 1 & 0 \\\hline
12) Connect OAuth2\cite{lib-connect-oauth} & 1 & 1 & 0 \\\hline
\multicolumn{1}{|c|}{\textbf{Total}} & \textbf{49} & \textbf{43} & \textbf{6} \\\hline

\end{tabular}
\caption{Evaluation results of \toolname on \datasetKnown.}
\label{tab:eval-ground-truth}
\vspace{-6mm}
\end{table}

\noindent
\textbf{Ground truth determination.} To determine the ground truth for \datasetUnknown, 
we perform both dynamic analysis and source code inspection with the OAuth flows simulated for each library. We deploy each library on a local server and implement client-side programs to simulate the OAuth flows according to the specification. In particular, for each property, we manually construct the parameters to initiate the corresponding OAuth requests and analyze the responses from the server. For example, to verify the result reported by \toolname for redirect URI property (P1), we first deploy the library on a local server and create an authorization server instance. We create two client instances--one with a benign redirect URI and another with a malicious one. We initiate an authorization request from the benign client, but replace the `\textit{redirect\_uri}' parameter with the redirect URI from the malicious client. We observe the traces generated by the server. If the server redirects the authorization response (with \textit{authorization code}) to the malicious URI, we mark a violation to the redirect URI property. Even if \toolname reports no property violation, we still investigate the library, in the same way to find if the library actually violates the property.

\begin{table*}[t]
\centering
\footnotesize
\begin{tabular}{|l|c|c|c|c|c|c|c|c|c|c|c|c|c|}
\hline
\textbf{OAuth-server libraries}
& \textbf{P1}
& \textbf{P2} 
& \textbf{P3} 
& \textbf{P4} 
& \textbf{P5} 
& \textbf{P6} 
& \textbf{P7} 
& \textbf{P8} 
& \textbf{P9} 
& \textbf{P10} 
& \textbf{\#Violation}
& \textbf{\#FP}
& \textbf{\#FN}\\ \hline

% Redirection properties
1) Node oauth2 server
& \checkmark 
& $\times$ 
& \checkmark 
& \checkmark 
& \checkmark 
& $\times$ 
& $\times$ 
& $\checkmark^*$
& $\times$ 
& $\times$ 
& 5
& 0
& 1\\ 
\hline

2) Oauth2orize
& \checkmark  
& $\times$ 
& $\times$ 
& \checkmark
& \checkmark 
& $\times$ 
& $\times$ 
& $\times^*$ 
& $\times$ 
& $\times$ 
& 6
& 1
& 0\\ 
\hline

3) Node oidc provider
& \checkmark
& $\times$
& \checkmark
& \checkmark
& $\checkmark^*$
& \checkmark
& \checkmark
& \checkmark
& $\checkmark^*$
& \checkmark
& 1
& 0
& 2\\ 
\hline

% removing this library from eval
% node-oauth20-provider
% & 
% & 
% & 
% &  
% & 
% & 
% &
% &
% &
% & 
% &  
% & \\ 
% \hline

4) Oauth2 server
& $\times$
& $\times$ 
& \checkmark
& $\times^*$ 
& \checkmark
& $\times$ 
& $\times$ 
& $\times$ 
& \checkmark
& \checkmark
& 5
& 1
&  0\\ 
\hline

5) Spring auth server
& \checkmark
& $\times$ 
& $\times$ 
& $\checkmark$
& $\checkmark$
& $\times$
& $\times$
& \checkmark 
& \checkmark
& \checkmark
& 4
& 0
&  0\\ 
\hline

6) Clouway server
& \checkmark
& $\times$ 
& $\times$ 
& $\times^*$ 
& \checkmark
& \checkmark
& \checkmark
& $\checkmark$
& $\times$ 
& $\times$ 
& 4
& 1
&  0\\ 
\hline

7) Jobmission server
& $\times^*$ 
& $\times$ 
& $\times$ 
& $\times$ 
& $\times$ 
& $\times$ 
& $\times$ 
& $\times^*$ 
& $\times$
& \checkmark
& 7
& 2
& 0\\ 
\hline

8) Apifest
& $\times$ 
& \checkmark
& \checkmark
& \checkmark
& $\times^*$ 
& $\times$ 
& $\times$ 
& $\checkmark$
& $\times$ 
& $\times$ 
& 5
& 1
&  0\\ 
\hline

9) Yoichiro server
& $\times^*$ 
& $\times$ 
& $\times$ 
& $\times$ 
& $\times$ 
& $\times$ 
& $\times$ 
& $\times$ 
& $\times$ 
& \checkmark
& 8 
& 1
&  0\\ 
\hline

10) OxAuth
& \checkmark
& $\times$ 
& \checkmark
& \checkmark
& $\times$ 
& \checkmark
& \checkmark
& \checkmark
& \checkmark
& \checkmark
& 2
& 0
&  0\\ 
\hline

\textbf{Total vulnerabilities:}
& 2
& 9
& 6
& 2
& 3
& 7
& 7
& 2
& 5
& 4
& 47
& 7
& 3\\ 
\hline

\end{tabular}

\caption{Evaluation results of \toolname on \datasetUnknown whose vulnerabilities are unknown. (\checkmark: library satisfies the property, $\times$: library violates the property). Here, $\times^*$ indicates a false positive (FP), where library satisfies the desired property, but \toolname marks it as a violation due to the limitation of the underlying static analysis. On the other hand, $\checkmark^*$ indicates false negative (FN), which could occur due to imprecisely resolved call-sites.}
\label{tab:property-check-result}
\vspace{-6mm}
\end{table*}

As presented in Table~\ref{tab:property-check-result}, we discover 47 confirmed vulnerabilities in the ten popular libraries of the authorization server, 24 of which were previously unknown. Security impacts of these vulnerabilities can cause severe attacks on the authorization server. We discuss these vulnerabilities and their implications below:\\
\noindent
\textbf{Redirect URI manipulation.} \toolname successfully identifies two libraries violating the property that the redirect URI from authorization request must match the URI registered by the client (P1). In addition, \toolname finds nine libraries violating the property that the redirect URI must be an \textit{absolute} URI (P2). For example, to check if the redirect URI parameter contains an absolute URI, the Node oauth2 server library matches it with an incorrect URI pattern ``[a-zA-Z][a-zA-Z0-9+.-]+:''. Violation of these properties allows the attackers to manipulate the redirect URI while initiating the authorization request and obtain a legitimate client's authorization code or access token directly from the authorization server. Our finding indicates that developers often do not follow the requirement for redirect URI, possibly because it may not seem security-critical from reading the specification.

\noindent
\textbf{Authorization code injection.} \toolname successfully identifies six libraries violating at least one property required to protect the clients against the authorization code injection attack. We find six libraries omitting or incorrectly implementing the property that the authorization code must be single-use (P3). This vulnerability allows the attackers to replay the authorization code intercepted from a legitimate client's request and obtain the access token. We also identify two libraries that do not bind the authorization code to any particular client (P4), allowing the attackers to exchange the code using a malicious client application. Additionally, \toolname successfully identifies four libraries that do not bind the authorization code to the redirect URI (P5), which prevents the authorization server from cross-checking the redirect URI with the one submitted at authorization endpoint. 
% Each of these vulnerabilities leaves an open door for the attackers to perform an authorization code injection attack that eventually allows the attackers to obtain a victim resource owner's information, which we describe with an attack example in Appendix~\ref{apx:attacks}. 
\toolname also finds seven libraries that do not support PKCE (P6 \& P7)--a recommended mechanism designed for the authorization server to mitigate authorization code injection attacks for public clients.\\
\noindent
\textbf{Access token injection.} \toolname successfully identifies five libraries vulnerable to the access token injection. All five libraries do not issue client-constrained access tokens from the token endpoint (P9) which limits the applicability of the token to a particular client. It prevents the attackers from using an access token that is obtained through a malicious client application. \toolname also identifies four libraries that store the access token in clear text format (P10). Storing access tokens as clear-text is vulnerable since it can expose the tokens to the attackers through accessing the local storage or launching a SQL injection attack.\\
\noindent
\textbf{CSRF attack.} \toolname successfully identifies two libraries that do not provide CSRF protection using OAuth's \textit{`state'} parameter (P8). By design, libraries supporting the PKCE are CSRF-protected. Therefore, libraries without PKCE must handle the authorization request's \textit{`state'} parameter to protect their clients against CSRF. \toolname reports a library as vulnerable if it neither supports PKCE, nor handles \textit{`state'} parameter.

\begin{table}[h]
\centering
\footnotesize
\begin{tabular}{|c|c|c|c|c|}

\hline
\multicolumn{1}{|p{1cm}|}{\centering \multirow{2}{*}{\textbf{Property}}} & \multicolumn{2}{p{3cm}|}{\centering \textbf{\oauthlint} ($\times$ slicing)\\Exec. time (min)} & 
\multicolumn{2}{p{3cm}|}{\centering \textbf{\toolname} ($\checkmark$ slicing)\\Exec. time (min)} \\ \cline{2-5}
& \datasetUnknown & \datasetKnown & \datasetUnknown & \datasetKnown \\ \hline
P6 & 48.3 & 64.8 & \textbf{2.1} & \textbf{3}\\\hline
P9 & 55.8 & 71.6 & \textbf{2.5} & \textbf{3.7}\\\hline
P10 & 34 & 52.9 & \textbf{1.7} & \textbf{2.2}\\\hline
\end{tabular}
\caption{Comparing \toolname against \oauthlint in terms of average  running time using the \datasetUnknown and \datasetKnown.}
\label{main-tab:eval-oauthlint}
\vspace{-6mm}
\end{table}

% \noindent
\subsubsection{Analysis Accuracy}~\label{subsec:accuracy} \toolname successfully identifies 47 property violations (Table~\ref{tab:property-check-result}) and reports 7 false-positive (FP) cases and 3 false-negative (FN) cases in \datasetUnknown, and 6 FN cases (Table~\ref{tab:eval-ground-truth}) in \datasetKnown. Our assessment finds the primary reason for the FP and FN cases is the spurious points-to targets from the WALA framework. Additionally, unsoundness can also occur for a few cases during Javascript library analysis. We summarize the reasons for the FP and FN cases in the following: 
\setlist{nolistsep}
\begin{itemize}[leftmargin=*]
    % Partially modeled library functions can cause false-positive cases during the analysis.
    \item Semantics of complex string operations (e.g., \texttt{subString} or \texttt{concate}) cannot be precisely modeled by WALA--which leads our analysis to be unable to keep track of the precise values held by some string variables.
    \item Semantics of functions for executing dynamically generated code (e.g., Javascript's $\mathtt{eval}$) are not currently modeled as we do not reason dynamic code in this work. 
    \item For reflected calls where a reference to a target method is computed at runtime (e.g., Java’s $\mathtt{invoke}$), our static analysis may not accurately resolve the target method, which can result in FNs.
    \item Javascript libraries with
    % use complex frameworks such as \textit{Node.js}~\cite{nodejs} which 
    dynamic features (e.g., dynamic module loading)  may cause WALA to generate a broken and imprecise callgraph that leads to FPs and FNs.
    % \item \tamjid{(need to update this)}We do not reason about the semantics of regular expressions. Instead, we over-approximate relevant regular expressions by checking 1) whether an API for performing sanitization using regular expressions is invoked, and 2) the arguments of the API are tainted by a redirect URI and absolute URI (P2 in Table~\ref{tab:list-of-properties}), respectively. 
    % % we currently check if the redirect URI string is matched against an absolute URI pattern~\cite{abs-uri-rfc} using the pattern matching APIs for the corresponding language.
    
\end{itemize}

%Also,  \redtext{\redsout{However, \toolname did not report any false-negative for property violations while analyzing the libraries, which demonstrates the correctness of our tool.}}\\

\subsubsection{Acknowledgments from Developers.}~\label{subsec:developers} We reported all the unknown vulnerabilities to the developers of the libraries. So far, we have received confirmation from the developers of eight libraries, who have acknowledged our findings. By the time of writing this paper, five libraries had fixed the redirect URI, PKCE, and authorization code issues as we reported; the other remaining issues are undergoing a review process. Developers of the two libraries responded that they are taking action on our reports, and the issues will be addressed in their next release. Developer of one library confirmed the issues we reported, but declined to take action as they no longer work for the corresponding vendor. Three classes of unknown vulnerabilities in these libraries led us to have three new CVE entries (CVE-2020-26877, CVE-2020-26937, and CVE-2020-26938). However, we have not yet received feedback from two libraries. We are making our best effort to reach the developers through different channels and help them fix the vulnerabilities.\\
\fbox{

\begin{minipage}{0.9\linewidth}

{\bf Result for RQ1:}
We identified 47 vulnerabilities in ten popular OAuth libraries, including developers' acknowledgment from eight libraries and three accepted CVE entries.

\end{minipage}

}\\

% \medskip

\begin{figure}[!t]

  \includegraphics[width=\linewidth]{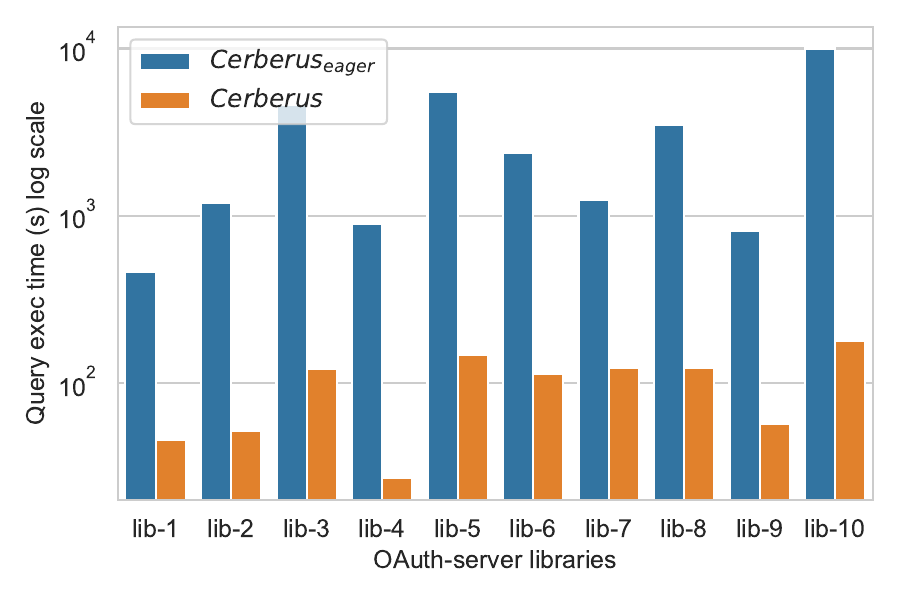}
% \vspace{-6mm}
  \caption{Performance comparison for libraries in \datasetUnknown (Table~\ref{tab:library-stats}) based on average query execution time (in \textbf{log scale}) between \toolname, which constructs the on-demand callgraph , and \toolvariant, which constructs the callgraph eagerly for the whole program. Bar(s) touching the maximum time ($10^4$) indicate the library not terminating or causing memory explosion within the maximum time limit (3 hours).}

  \label{fig:performance}
% \vspace{-6mm}
\end{figure}

\subsubsection{Performance Evaluation}~\label{sec:ablation}

To answer \textbf{RQ2} and evaluate the effectiveness of our query-driven algorithm discussed in Sec.~\ref{sec:sdg}, we compare \toolname with its variant, which constructs the callgraph for the entire program eagerly (noted as \toolvariant). Unlike \toolname, which constructs an on-demand callgraph based on the OAuth query and prunes away any program components that are irrelevant with respect to the query, \toolvariant does not consider any relevance with the query. Instead, it constructs the callgraph eagerly for the whole program. We execute our queries with both \toolvariant and \toolname and compare their performances.

% One of the design choices for \toolname is to construct the SDG from a context-insensitive callgraph which is constructed eagerly during the query execution. Particularly, in this approach, a query is executed over the entire SDG of the input program, regardless of their relevance to OAuth flow. Another design choice is to construct a query-driven on-demand SDG which utilizes the product automaton based pruning while constructing the callgraph for the input program. To answer \textbf{RQ2}, we investigate the performance of two different design choices for \toolname: (1) \toolname$^\dagger$, which constructs callgraph eagerly (without pruning) for the whole-program, (2) \toolname, which constructs callgraph on-demand (with pruning). 

Fig.~\ref{fig:performance} shows the average query execution time for the two different design choices of callgraph construction by our tool: \toolname and \toolvariant. Y-axis shows the average time in seconds (in log scale) to answer queries for each library (X-axis) selected for the evaluation. Our analysis shows that our tool, \toolname, which executes queries over an on-demand callgraph, is on average $25\times$ faster than \toolvariant, which eagerly constructs callgraph. For example,\textit{ Spring Authorization Server} and \textit{Node Oidc Provider} library take more than one hour to answer an OAuth query using \toolvariant. On the other hand, the same query is answered by 147 and 121 seconds, respectively, using the on-demand approach by \toolname. This significant query execution time with \toolvariant is because, by design, the analysis also visits the program components that are not directly relevant to the OAuth protocol (e.g., database models and operations). On the other hand, \toolname only visits the program components relevant to the submitted OAuth query and therefore requires significantly less time to answer a given query. 

However, we limit the execution time to 3 hours for executing the queries. In other words, the query execution automatically terminates if it can not find any result by 3 hours. We find that one of the large libraries, \textit{OxAuth} ran into a memory explosion after running for more than 2 hours using \toolvariant. However, it takes 178 seconds to answer the same queries using \toolname, which shows the effectiveness of our tool for analyzing large-scale implementations. 

\textbf{Comparison with existing tools.}
% To the best of our knowledge, there is no existing work that performs (whitebox) analysis on the implementation of OAuth-server (i.e., service provider) applications. However, there exists tools that analyze OAuth properties by analyzing the flows observed from client-side applications, while treating the server-side as blackbox. Among these analysis tools, OAuthLint~\cite{DBLP:conf/kbse/RahatFT19} is the closest to our work as they also statically analyze source code of client applications to check OAuth properties.
We further extend our effort to compare \toolname against \oauthlint, which only supports OAuth properties with data-flow predicates. We managed to encode only three of our properties (out of ten) using their predicates with our best effort. 
% In addition, \oauthlint runs its analysis over the entire client application of OAuth. 
As shown in Table~\ref{main-tab:eval-oauthlint}, for the three properties supported by both tools, \toolname is $22\times$ and $21\times$ faster than \oauthlint on \datasetKnown and \datasetUnknown, respectively.

% However, as shown in Figure~\ref{fig:performance}, when executing the queries with \toolvariant, 3 libraries -- \textit{node-oidc-provider}, \textit{apifest-oauth20} and \textit{oxAuth} could not return any result within the limit of 10 minutes. To further investigate their performance, we increase the maximum time limit to 1 hour. With this extended setting, one library (\textit{apifest-oauth20}) returned the result after running for 22 minutes, and two other libraries ran into memory explosion after running for 27 and 32 minutes, respectively. However, when we used \toolname for these three libraries, they successfully returned the results by 121, 123, and 178 seconds, respectively, which shows significant performance improvement by \toolname when compared to the \toolvariant.

\medskip
\fbox{
\begin{minipage}{0.9\linewidth}
{\bf Result for RQ2:}
Our query-driven analysis is $25\times$ faster in checking OAuth security properties than the eager analysis.
\end{minipage}
} 
% \subsubsection{Applications Using OAuth Libraries}

% \redtext{RR8: Provide more details on the extensibility of Cerberus with a running example.}

\subsubsection{Extensibility}~\label{subsec:extensibility}
% There have significant efforts in recent years to improve the security of OAuth--which led to new proposals of OAuth extensions.
To show the extensibility of \toolname, we use a recent property from DPoP (Demonstrating Proof-of-Possession)~\cite{dpop-ietf}, which is a new draft describing a mechanism for sender-constrained OAuth token. When sending a token request, the client sends an additional DPoP token in JWT format~\cite{jwt} that allows the server to verify the proof of possession of the token. The server first verifies the token using the key embedded within the token. If verified, it adds the thumbprint of the key with the access token. Otherwise, the token request is rejected with an error response. One can use \toolname's predicates to express this query as follows:

{\footnotesize
\begin{verbatim}
1. P :- invoke(L1, token_req), OAuthTag(L2, dpop),  
2.      OAuthTag(L3, key), invoke(L4, jwt_verify), 
3.      OAuthTag(L5, access_token), invoke(L6, add_thumbprint), 
4.      error(L7, _), branch(L8,X,L6,L7), followBy(L1,L2),
5.      followBy(L1,L5), depOn(L2,L4),depOn(L3,L4),
6.      depOn(L4,X), depOn(L5,L6), depOn(L3,L6).
\end{verbatim}
}
As shown from the above query, it is straightforward for a domain expert to use \toolname's built-in predicates to define new properties. Specifically, given a new property, a domain expert only needs to encode the new domain-specific information using predicates \texttt{OAuthTag} and \texttt{invoke}. E.g., for the DPoP property, the extra taint sources will be  \texttt{dpop}, \texttt{key} and \texttt{jwt\_verify}.

\section{Related work}~\label{sec:related}
Since OAuth is a widely used and security-critical multi-party protocol, many researchers have studied OAuth at the protocol and implementation level for various platforms such as mobile~\cite{wang2015vulnerability}, web~\cite{sun2012devil,cao2014protecting}, and IoT~\cite{sciancalepore2017oauth,oauth-cirani}. Yang et al.~\cite{oauth-yang} study the attacker models to perform the common OAuth attacks (e.g., impersonation attacks, CSRF attacks) in web applications. Chen et al.~\cite{chen2014oauth} present the first in-depth study on OAuth attack vectors caused by the mistakes made by developers in client applications. However, manual analysis on client applications cannot be used to check logic flaws in \oauthserver. Although, the manual approach can be effective for checking particular flaws, it is not scalable and might miss vulnerabilities. In comparison, \toolname is the first automated and scalable tool to systematically check the logical flaws on \oauthserver implementation. In addition, \toolname defines new security properties and detects novel vulnerabilities in real-world \oauthserver. 
% Even if the security analysts can spend a large amount of manual efforts to check following the method in the paper~\cite{chen2014oauth}, they won't be able to check six of out ten properties we proposed (P3,4,5,6,7,10). This is because Chen et al.'s work focus on analyzing mobile apps. Thus, their manual analysis techniques of the mobile apps cannot be used to check the logic flow between the relying party servers and service provider servers. 
Emerson et al.~\cite{oauth-emerson} propose an OAuth-based central access management system to provide a secure authentication scheme for IoT devices. Calzavara et al.~\cite{calzavara2018wpse} study the browser-side (i.e., client-side) security for using OAuth while considering the \oauthserver as black-box. Veronese et al.~\cite{veronese2020bulwark} propose a network-traffic-based security monitoring system for different entities of OAuth. However, these studies are focused on the security implications with respect to the client-side flow and depend on manual analysis by security experts such as monitoring the network traffic or inferring the protocol flows and cannot be applied or extended to detect missing or incorrect security checks on the \oauthserver implementation.

Researchers also study automated analysis to find security issues in client-side OAuth flow. Yang et al.~\cite{yang2018vetting} develop a symbolic execution-based testing tool to check the correctness of OAuth SDKs that provides APIs for client applications. Rahat et al.~\cite{DBLP:conf/kbse/RahatFT19} build a static taint analysis tool to detect five categories of OAuth-specific data flow in Android applications. While these tools are effective for checking flaws on client applications, as we show in our evaluation, they are not designed to analyze the large scale codebase on \oauthserver. Instead of analyzing the entire application, our tool is designed to automatically extract and reason about the OAuth-specific partial programs to identify security critical logical mistakes.

% In recent years, researchers have been studying the challenges of detecting incorrect API usage in different security domains. He et al. implemented SSLint~\cite{he2015vetting} that uses static analysis and graph signature matching to identify incorrect API usage for SSL. Egele et al.~\cite{egele2013empirical} studied cryptographic API misuse and used static program slicing to detect incorrect cryptographic operations in Android applications. These works focus on high-level API usage patterns and do not scale well for large programs, whereas our tool is designed to identify fine-grained complex properties in a scalable fashion. 
Graph-based query approaches~\cite{martin2005finding, goldsmith2005relational,rahat2018maximizing} have been applied to solve a wide variety of security issues such as API misuse detection~\cite{zhang2014semantics,aafer2018precise,staicu2018synode,sven2019investigating,rahaman2019cryptoguard,lv2020rtfm}, tainted data-flow analysis~\cite{yang2012leakminer,li2014know,yamaguchi2015automatic,cheng2018dtaint,fass2021doublex} and threat detection~\cite{gao2021system,park2021identifying}. He et al. implemented SSLint~\cite{he2015vetting} that uses static analysis and graph signature matching to identify incorrect API usage for SSL. Egele et al.~\cite{egele2013empirical} studied cryptographic API misuse and used static program slicing to detect incorrect cryptographic operations in Android applications. These works focus on high-level API usage patterns, whereas \toolname is designed to identify fine-grained properties (e.g., expressions). Graph traversal using different graph representations\cite{aho2007compilers, ferrante1987program, yamaguchi2012generalized} of source code has also been used for checking code properties in various applications~\cite{fass2019hidenoseek,shoshitaishvili2015firmalice,cao2015edgeminer}. In addition to these classic representations, combined graphs~\cite{limining} have also been used for vulnerability detection. For example, Yamaguchi et al.~\cite{yamaguchi2014modeling} use a combined graph representation called Code Property Graph to discover vulnerabilities. However, though these implementations are effective for checking common vulnerabilities (e.g., buffer overflows) in specific languages (e.g., C/C++), we cannot evaluate or compare against them as they do not support the languages (Java and Javascript) we target in this paper. In addition, these works eagerly construct the graphs for the entire program, which, as our evaluation demonstrates (Fig.~\ref{fig:performance}), does not scale for large-scale programs such as \oauthserver implementation.
\section{Discussion}~\label{sec:discussion}
While designing \toolname, we strived to achieve a good trade-off between expressiveness and scalability. For complex semantics (e.g., secure computation, storage modeling, etc.) that are difficult to model precisely, our current static analysis leverages data- and control-dependencies to over-approximate the actual semantics, which, in theory, can lead to spurious execution paths. However, as discussed in Sec.~\ref{sec:eval}, our tool achieves a low false positive rate despite our modeling.
% during the runtime and result in FP and FN cases. 
% Also, as discussed in Section~\ref{sec:eval}, due to the dynamic nature of some Javascript frameworks (e.g., \textit{Node.js}~\cite{nodejs}) used by the OAuth libraries, WALA can generate a dense callgraph that can lead to FP and FN cases.
%\yuan{To handle the dynamic features (e.g., regex matching) of some OAuth properties, we design a hybrid analysis component by adding lightweight delta testing for the relevant dynamic program component (see details in Section~\ref{sec:hybrid}). We show that the hybrid analysis help to future reduce our FP cases from 9 to 7. }
Secondly, dynamic features of Java and Javascript, such as reflective calls, dynamic class loading, and exceptional handling can result in false negatives. Our current implementation of hybrid analysis currently provides limited support to handle dynamic features (e.g., reflective calls with string constants). 
% We plan to further mitigate this issue by integrating  Tamiflex~\cite{tamiflex} and ACG~\cite{acg} tools for systematically reasoning about dynamic features into the \toolname toolchain.

% One future work is to further reduce the FP \tamjid{and FN} cases of \toolname.
% Many of these FP and FN cases come from the imprecisely resolved call-sites by WALA, which is a well-known limitation of static analyzers. In addition, currently we  do not reason about the semantics regular expression or dynamically generated code--which can also lead to FP and FN cases. As a result, it would be interesting to explore integrating light-weight dynamic analysis and develop precise abstractions for the semantics of complex functions to improve the precision of the analysis. Additionally, reasoning the semantics of dynamically generated code can further improve the precision of \toolname. 
% Since the semantics of dynamic code depends on the logic of the dynamic code generator, the analysis needs to reason about the dependencies and data flow between the dynamic code generator and generated code to be able to determine the behavior of the dynamically generated code.

Our study in this paper focuses on OAuth 2.0, a prevalent multi-party protocol for authorization. \toolname is designed based on the standard OAuth specification~\cite{rfc6749}. Therefore, any implementation that follows the standard specification can be analyzed using our proposed approach, which exhibits the generality of our work. Additionally, \toolname can be extended to other relevant protocols like OpenID Connect~\cite{openid-spec}. Since OpenID Connect uses similar grants and flows as OAuth 2.0, the vulnerabilities we address in this paper are also applicable for OpenID Connect supported servers. In particular, \toolname's predicates can be extended to check OpenID properties using our query-based method. However, as our current analysis does not model the cryptographic APIs (e.g., RSA verification), some OpenID Connect flows (e.g., hybrid flow) that involve cryptographic operations cannot be checked by our tool. %In addition, though our security properties are focused on two primary endpoints of OAuth, developers can use \toolname to check properties for other endpoints (e.g., token introspection endpoint) or any custom endpoint by specifying them as regular expression. 
%We leave the security properties relevant to the usage of other OAuth-based protocols such as OpenID Connect or Financial-grade API (FAPI)~\cite{fapi} as future work.
Additionally, it would be an interesting future work to apply \toolname on a large scale of OAuth service provider implementations to check the prevalence of the issues we identified. The results might be similar because we study very popular libraries for OAuth service provider implementations, and developers usually just call these libraries' APIs instead of building their own implementations. Even worse, if developers start from scratch to build their service providers, they might make more security mistakes. In addition to OAuth, our proposed method can also be applied to cross-check applications against policy regulations such as GDPR~\cite{rahat2021automated} and CCPA~\cite{goldman2020introduction}.

Finally, our analysis in this paper is focused on the attacks that utilize incorrect or logical implementation mistakes made during the OAuth flow. Our tool is designed to find violations in the implementation with respect to the queries representing the violation pattern. While our properties cover the OAuth-specific attacks commonly observed on the OAuth servers, attackers might exploit other vulnerabilities to attack the server. For example,  Attackers might leverage generic web/mobile vulnerabilities and perform web-based/mobile attacks (e.g., SQL-injection) to steal OAuth credentials or protected resources of the resource owner. Detecting those generic vulnerabilities is beyond the scope of this paper.

\section{Conclusion}~\label{sec:concl}
In this paper, we have presented \toolname, an automated analyzer that can discover logic vulnerabilities in \oauthserver libraries that service providers widely use. To efficiently detect OAuth violations in large codebases, \toolname employs a query-driven algorithm for answering queries about security-critical OAuth properties. 
To demonstrate the effectiveness of \toolname, we evaluate it on datasets of popular
\oauthserver libraries with millions of downloads. Among these high-profile libraries, \toolname has discovered 47 vulnerabilities from ten classes
of logic flaws, 24 of which were previously unknown and led to new CVEs.% entries.
% \redtext{RR9: Address reviewers' comments related to writing and presentation of results.}

\section{Acknowledgements}
We are grateful to the anonymous reviewers for their insightful and constructive feedback and suggestions. This work is supported in part by \textit{National
Science Foundation} under the award numbers 1943100, 1920462, 2114074, and 1908494, by \textit{DARPA} under the agreement number N66001-22-2-4037, by \textit{Google Faculty Research}, and \textit{Facebook Faculty Fellowship} awards. The views and conclusions contained in this document are those of the authors and should not be interpreted as representing the official policies, either expressed or implied, of the funding agencies.

\bibliographystyle{ACM-Reference-Format}
\balance
\bibliography{main}

% \input{appendix}

% \input{summary-changes}

\end{document}